\documentclass[preprint]{aastex}
\usepackage{graphicx}
\usepackage{xcolor}

\usepackage{natbib}

\newcommand{\be}{\begin{equation}}
\newcommand{\ee}{\end{equation}}
\newcommand{\nn}{\mbox{} \nonumber \\ \mbox{} }
\newcommand{\ba}{\begin{eqnarray}}
\newcommand{\ea}{\end{eqnarray}}
\newcommand{\om}{\omega}

\newcommand{\Bf}{{magnetic field}}
\newcommand{\Ef}{{electric field}}
\newcommand{\Bfs}{{magnetic fields}}

\newcommand{\mss}{magnetospheres}

\newcommand{\Lf} {{Lorentz factor}}

\newcommand\eg{{\it{e.g.}}}

\newcommand\ie{ \it{i.e.\ }}

\newcommand\lo{\mathrel{\raise.3ex\hbox{$<$}\mkern-14mu\lower0.6ex\hbox{$\sim$}}}
\newcommand\go{\mathrel{\raise.3ex\hbox{$>$}\mkern-14mu\lower0.6ex\hbox{$\sim$}}}

\begin{document}
\title{Interpreting Crab Nebula's Synchrotron Spectrum: Two Acceleration Mechanisms
}

\author{Maxim Lyutikov$^1$, Tea Temim$^2$, Sergey Komissarov $^3$, Patrick Slane $^4$,  Lorenzo Sironi $^5$,  Luca Comisso $^5$}
\affil{$^1$
Department of Physics and Astronomy, Purdue University, 
 525 Northwestern Avenue,
West Lafayette, IN, USA
47907\\
$^2$ Space Telescope Science Institute, 3700 San Martin Drive, Baltimore, MD 21218, USA\\
$^3$  School of Mathematics,
University of Leeds, LS29JT
Leeds, UK\\
$^4$Harvard-Smithsonian Center for Astrophysics, 60 Garden St., Cambridge, MA 02138,
USA\\
$^5$Department of Astronomy, Columbia University, 550 W 120th St, New York, NY 10027, USA
}

\begin{abstract}
We outline a model of the Crab Pulsar Wind Nebula with two different
populations of synchrotron emitting particles, arising from two
different acceleration mechanisms: (i) Component-I due to Fermi-I
acceleration at the  equatorial portion of the
termination shock, with particle spectral index $p_I\approx 2.2$ above the injection break  corresponding to $\gamma_{wind} \sigma _{wind} \sim 10^5$, peaking in the UV
($\gamma_{wind} \sim 10^2$ is the bulk \Lf\ of the wind, $\sigma _{wind} \sim 10^3$ is wind magnetization);
(ii)  Component-II due  to acceleration at  reconnection layers in the bulk of the turbulent Nebula, with particle index $p_{II} \approx 1.6$. The model requires relatively slow but highly magnetized  wind. For both
components the overall cooling break is in the infra-red at $\sim 0.01$ eV, so that 
  the  Component-I is  in the fast
cooling regime (cooling frequency below the peak frequency).   In the optical band Component-I produces emission with the
cooling spectral index of $\alpha_o \approx 0.5$, softening towards the edges due to radiative losses.
 Above the cooling break, in the optical, 
UV and X-rays,  Component-I  mostly overwhelms  Component-II. We hypothesize  that acceleration at large-scale current sheets in the turbulent nebula (Component-II) extends to the synchrotron burn-off limit
of $\epsilon_s \sim$  100 MeV. Thus in our model acceleration in   turbulent reconnection (Component-II)
can produce both hard radio spectra and occasional gamma-ray flares. 
This model may be applicable to a broader class of high energy astrophysical objects, like AGNe and GRB jets, where often radio electrons form a different population from the high energy electrons.
\end{abstract}

\section{Introduction}
The Crab Nebula is the paragon of the high energy astrophysical
source. Understanding physical processes operating in the Crab has
enormous implications for high energy astrophysics in general.
Conventionally, the particle acceleration was assumed to occur at
the wind termination shocks, presumably via the Fermi-I mechanism
\citep{reesgunn,1984ApJ...283..694K,1984ApJ...283..710K}.  The expected  particle spectral index  $p=2.2-2.4$ \citep[\eg][]{1987PhR...154....1B} matches nicely the inferred non-thermal  X-ray synchrotron spectrum, see \S \ref{overal} and Fig. \ref{Crab-spectrum-pictures-1}. (We adopt the notations for the spectral energy distribution
as $F_\nu \propto \nu^{-\alpha }$, and particle energy distribution
$f(\gamma) \propto \gamma^{-p}$.)

A strong argument in favor of particle acceleration at the (equatorial
part of the) termination shock  comes from the fact that the low
magnetization numerical models of the Crab Nebula
\citep{2004MNRAS.349..779K,DelZanna04} are able to reproduce the
morphological details of the Crab Nebula down to intricate details.
Until recently, acceleration at the termination shock was the dominant
paradigm \citep{reesgunn,1984ApJ...283..694K,1984ApJ...283..710K}. Consistent with this picture is the fact that  low-sigma shocks are efficient particle accelerators \citep{2011ApJ...741...39S,2013ApJ...771...54S}.

One of the biggest problems, recognized by \cite{1984ApJ...283..710K},
was the radio emission of the Crab and other pulsar wind nebulae (PWNe).
First, the radio spectra of PWNe are very hard, with the spectral
index $\alpha$ approaching zero in some cases
\citep[][]{2001SSRv...99..177R,2014BASI...42...47G,2017SSRv..207..175R}.
The implied particle spectral index $p$ is then approaching $ p \rightarrow
1$.  The spectral index $p< 2$ is
not expected in Fermi-I acceleration models that typically have
$p\geq 2$ \citep[\eg][]{1987PhR...154....1B}. Secondly, the number
of radio emitting particles is much larger than what is expected
in magnetospheric pair production
\citep{1984ApJ...283..710K,1999A&A...346L..49A,AronsScharlemann,2010MNRAS.408.2092T},
see also \S \ref{multiplicity}. (We suggest a way to obtain a hard
spectrum, but the multiplicity problem remains.)

Another important observation that constrained the structure of PWNe
was the discovery of flares from the Crab Nebula
\citep{Tavani11,2011Sci...331..739A,2013ApJ...765...56W}, which further
challenged our understanding of particle acceleration in the PWNe.
The unusually short durations, high luminosities, and high photon
energies of the Crab Nebula gamma-ray flares require reconsideration
of our basic assumptions about the physical processes responsible
for acceleration of (some) high-energy emitting particles in the
Crab Nebula, and, possibly in other high-energy astrophysical
sources.

Even before the discovery of Crab flares, \cite{2010MNRAS.405.1809L}
argued that the observed cutoff in the synchrotron spectrum of the Crab
Nebula at $\sim 100$ MeV in the persistent Crab emission
is too high to be consistent with the diffusive shock acceleration. Indeed, balancing
electrostatic acceleration in a regular electric field with synchrotron
energy losses yields a maximum synchrotron photon energy
\citep{1996ApJ...457..253D,2010MNRAS.405.1809L} 
\be \epsilon_{\rm
max} \sim \eta \hbar { m c^3 \over e^2} \approx 100 \, \eta \mbox{ MeV}.
\label{emax} 
\ee
 where $\eta$ is the ratio of electric to magnetic
field strengths. Since the high conductivity of astrophysical plasmas
ensures that in most circumstances $\eta<1$, the observed value of
the cutoff is right at the very limit. During the gamma-ray flares
the cutoff energy approached even higher value of $\sim 400 {\rm\ MeV}$,
suggesting a different acceleration mechanism
\citep[][]{2012MNRAS.426.1374C,2012ApJ...746..148C,2014PhPl...21e6501C,2013ApJ...770..147C,2017JPlPh..83f6301L,2017JPlPh..83f6302L,2018JPlPh..84b6301L}
- particle acceleration during reconnection events in highly
magnetized  plasma. In particular,
\cite{2017JPlPh..83f6302L,2017JPlPh..83f6301L} developed a model
of explosive reconnection events, whereby particles are accelerated
by a charged-starved \Ef\ during the X-point collapse in
relativistically magnetized plasma.

In this paper we argue that there are two acceleration mechanisms
in PWNe  (which we call, naturally, Component-I and Component-II): Component-I is due to Fermi-I acceleration at the termination shock  (and just thermalization at some parts of the shock)  and Component-II is due to plasma turbulence with self-consistent generation of reconnecting current sheets in the bulk of the Nebula.  \citep[Therefore, this scenario is different from the models that invoke turbulence with a pre-existing current sheet;][]{1999ApJ...517..700L}. The possibility of having
two acceleration mechanisms in PWNe has been suggested by
\cite{1984ApJ...283..710K} and then further discussed by
\cite{1996MNRAS.278..525A,2002A&A...386.1044B,2010A&A...523A...2M,2013MNRAS.433.3325S,2014MNRAS.438..278P}.
Of particular relevance are the works of \cite{2014MNRAS.438.1518O,2015MNRAS.449.3149O} who discussed two particle populations and argued that X-ray emission is produced from particles injected in the equatorial
regions, while radio electrons are accelerated in the bulk. These are the underlying assumptions in the present study.
Also, \cite{2013MNRAS.428.2459K} suggested that magnetic dissipation in the bulk of the Nebula accelerates particles.

Here we discuss  a more detailed model of the two component structure
of the Crab Nebula, concentrating on  the
broadband SED and the spatial distribution of the emission in different
bands.

\section{Overall spectrum: observations}
\label{overal}

\subsection{Mean features of the non-thermal component}
\label{breaks}
A number of authors have made fits to the spectral energy distribution
of the Crab Nebula \citep[see, \eg\ reviews
by][]{2017ASSL..446..101Z,2017SSRv..207..175R}.
Let us briefly summarize them.
 (see Fig. \ref{Crab-spectrum-pictures-1}):

\begin{figure}[h!]
\includegraphics[width=.8\columnwidth]{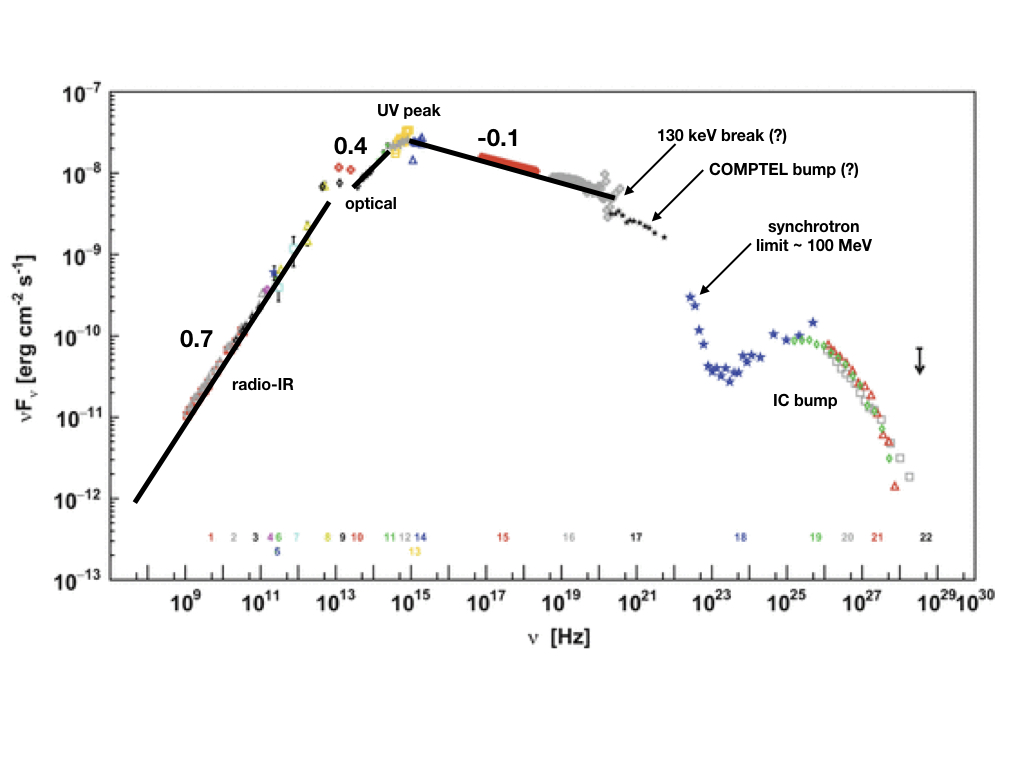}
\caption{
Broad-band spectrum of the Crab Nebula, reprinted from
\protect\citep[from][]{2017ASSL..446..101Z}. The infrared points shown  are contaminated by dust emission. We highlighted some
features of the spectrum; numbers indicate scaling of $\nu F_\nu$
with frequency. See text for details.
}
\label{Crab-spectrum-pictures-1}
\end{figure}

\begin{enumerate}

\item In the radio the spectral index is $\alpha_r =0.3$ (so that
$ \nu F_\nu \propto \nu^{0.7 }$). It is homogeneous over the Nebula
\citep{bietenholz_97}.

\item In the infra-red the situation is complicated -- a large thermal component from filaments complicates the analysis, see \S \ref{SpectralmapsinIR}

\item In the optical the spectrum softens to $\alpha_o \sim 0.6$ (so
that $ \nu F_\nu \propto \nu^{0.4 }$).

\item The overall spectral distribution of $\nu F_\nu$ in the Crab
Nebula has a peak in the UV, at $\sim 1$ eV. Above the peak the
spectrum has $\alpha \approx 1.1 $

\item An additional feature present in the hard X-ray spectrum is
softening at $\sim 130$  keV, with a corresponding break with
$\Delta \alpha = 0.43$ \citep{2010A&A...523A...2M} (see \S \ref{COMPTEL}
for further discussions)

\item There is also a claim of a mild spectral bump at above $\sim
1$ MeV \citep{1998A&A...330..321V}, 
though this may be due to the low statistics of the early COMPTEL
data set (Kuiper, priv. comm.; also \S \ref{COMPTEL}).


\item There is an exponential cut-off around $100$ MeV
\citep{2010ApJ...708.1254A}.

\item A new component appears above 1 GeV, peaking at around 100 GeV
\citep{2010ApJ...708.1254A,2006A&A...457..899A}

\end {enumerate}

\subsection{The COMPTEL break and a bump}
\label{COMPTEL}
There are possibly important features in the high energy Crab Nebula
spectrum that seem to be not given enough attention previously.
First, there is a softening at $\sim 130$  keV, with a break with
$\Delta \alpha = 0.43$ 
\citep{2010A&A...523A...2M}. It is not clear if this feature is
actually physically present. In \cite{2001A&A...378..918K} (their
Fig. A.1) there seems to be a factor of $1.5-2$ jump between the
continuation of $\sim$ 100  keV spectrum and COMPTEL points above 1
MeV. This could be due to cross-detector uncertainty (Lucien Kuiper,
priv. comm.). Yet a distinct difference in spectral slopes between
$\sim 1 - 40$~MeV 
\citep[][]{1998A&A...330..321V} may indicate a
different emission component.  The  data from SPI on
INTEGRAL \citep{2009ApJ...704...17J} 
indicate that a broken power law spectrum is statistically preferred over a single power law 
\citep[see Fig. 1
in][]{2009ApJ...704...17J}.


Thus, in our model the  $\sim 100$  keV - MeV break and the bump arise from a superposition of
two components, \S \ref{bump1}. If confirmed, the    COMPTEL break and a bump can be an interesting test
of the model. We encourage
further analysis of the Crab Nebula data in the MeV range.

\section{Internal structure - observations}
\label{internal}
Above, we discussed the overall properties of the Crab Nebula. Next
we highlight important observations related to the internal structure
of the Crab Nebula.

\subsection{X-rays}
\label{Xrays}

X-ray images of the Crab Nebula show interesting morphological variations.
 At  lower energies, $0.5-1.2$ keV, the Nebula is considerably more extended than in higher energies.  Overall, this is consistent with cooling, though
 some features (\eg\ southward directed elongated feature) are both much brighter in
lower energies, and have a corresponding features in radio and IR.

Spectral maps of the Crab in X-rays above 10~ keV,  {\ie} above the Chandra band obtained with  NuStar, are quite different
\citep[see Fig. 8 of][]{2015ApJ...801...66M,2017SSRv..207..175R}. The  map of spectral indices
indicates that  above  10~ keV the hardest part is off the main torus, and hardens towards
the  SE (South-East). This contrasts with the fact that  the Chandra images
show that  the high energy part (up to 7 keV) is nearly all torus. (The effect is not likely due to different spatial resolutions at different energies.)

The morphological map of the X-ray spectral index
shows interesting  structure: above 10~keV the hardest part is off the main torus, towards the low-left jet \citep{2015ApJ...801...66M}. The torus spectrum softens sooner (closer to the center) than the SE part.
This can be interpreted as due to the appearance of a new component, \S \ref{Xrays}.

We interpret the X-ray intensity and spectral maps as follows. The $\sim 0.1-10$  keV
intensity is dominated by Component-I; the increasing size of the emission region towards
softer energies is due to radiative cooling (X-ray particles have cooling times of
the order of a year). Above 10~ keV we start to see Component-II, that eventually
produces the COMPTEL bump, \S \ref{COMPTEL}. 

\subsection{Spectral maps in optical: softening towards the edges}

The map of the spectral index at optical wavelengths  \citep{1993A&A...270..370V}
 shows values of $\alpha$ as low as 0.57 in the vicinity of the pulsar with gradual steepening towards the outer edges of the Nebula to values just above unity. The clear softening of the spectrum towards the outer nebula implies that the optical-emitting electrons have cooling times shorter than the age of the Nebula. 

\subsection{The infrared}
\label{SpectralmapsinIR}

\begin{figure*}[h!]
\center
\includegraphics[width=.9\columnwidth]{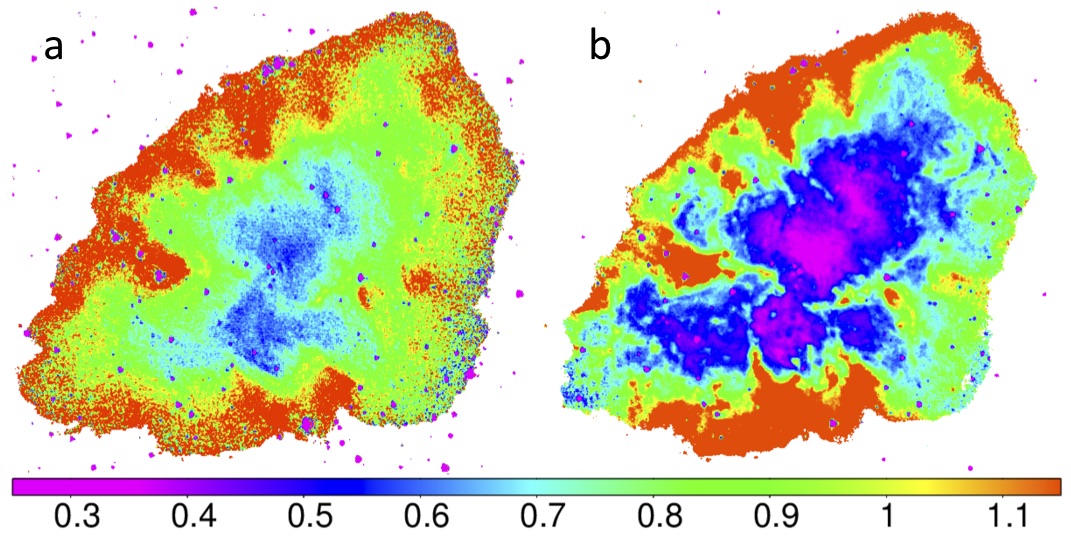} 
\caption{
Left Panel: Map of the spectral index calculated as a difference
between the synchrotron-dominated $3.6 \mu$m and $4.5\mu$m images. The torus structure is evident in blue, with gradual steeping towards the outer nebula. The uncertainty on the absolute value of the spectral index is estimated to be $\pm0.3$.
Right Panel: Map of the spectral index calculated as a difference
between $3.6\mu$m and $8.0\mu$m images. While the presence of the ejecta filaments at 8.0~\micron\ make the spectral index in the outer nebula unreliable, the spectral index for the inner torus seen in magenta is $0.3\pm0.1$. This matches the spectral index of the radio electrons. The color bars indicate the value of the spectral index.
}
\label{alpha-map}
\end{figure*}

As a part of new analysis, we reprocessed the \textit{Spitzer}
Infrared Array Camera (IRAC) data originally presented in
\cite{2006AJ....132.1610T} using updated calibration files, extended
source corrections, and the final v18.5 of the MOPEX (MOsaicker and
Point source EXtractor) software. We used the IRAC 3.5, 4.5, and
8.0~\micron\ images 
from \cite{2006AJ....132.1610T}
to create spectral index maps. The images
 were
background subtracted using background levels of 0.18. 0.18, and
8.72 MJy/sr, respectively.  The IRAC surface brightness correction
for extended sources was applied to the images (0.91, 0.94, and
0.74, respectively). The images were also corrected for extinction
using $A_{\lambda}/A_{K}$ values of \cite{2005ApJ...619..931I} and
an $A_{K}$ value of 0.194, derived from the hydrogen column density
towards the Crab Nebula of $N_H=3.54\times10^{21}\:cm^{-2}$
\citep{2001A&A...365L.212W} and the relation $N_{H}/A_{K} =
1.821\times10^{22}{\rm\ cm}^{-2}$ \citep{1989ESASP.290...93D}. This led
to extinction correction factors of 1.1 for the 3.6~\micron\ and
1.08 for 4.5~\micron\ and 8.0~\micron images. Finally, we used the nominal
wavelengths of 3.550~\micron, 4.493~\micron, and 7.872~\micron\ to
calculate the spectral index maps.

The two IR spectral index maps are shown in Fig.~\ref{alpha-map}. The left panel show the map calculated from the  IRAC 3.6 and 4.5 \micron\ images that are dominated by synchrotron emission. The torus structure, evident in blue, has a spectral index of 0.55. The index steepens with distance from the torus, consistent with previous results. While the relative values of the spectral indices have small uncertainties, the uncertainty on the absolute value of the index for the map on the left is $\pm0.3$, assuming a conservative estimate for the IRAC flux calibration of $\sim$5\%. We therefore produced a second spectral index map using the IRAC 3.6 and 8.0 \micron\ images, which provide larger wavelength spacing and reduce the absolute uncertainty to $\pm0.1$. This map is shown in the right panel of Fig.~\ref{alpha-map}. Since the IRAC 8.0~\micron\ image has a large contribution from the ejecta filaments, the spectral index is significantly modified in the outer nebula. We can, however, use this map to more reliably constrain the absolute value of the spectral index for the inner torus, which clearly stand out in magenta. Here, the torus has a spectral index of $0.3\pm0.1$. This is consistent with the continuation of the radio spectrum (which is clearly in the regime where radiative
losses are not important).  Similar spectral indices were found by
\cite{2002ASPC..271..161G} when comparing $16 \mu$m data with ISO
from $2 \mu$m.

\begin{figure}[h!]
\includegraphics[width=.49\columnwidth]{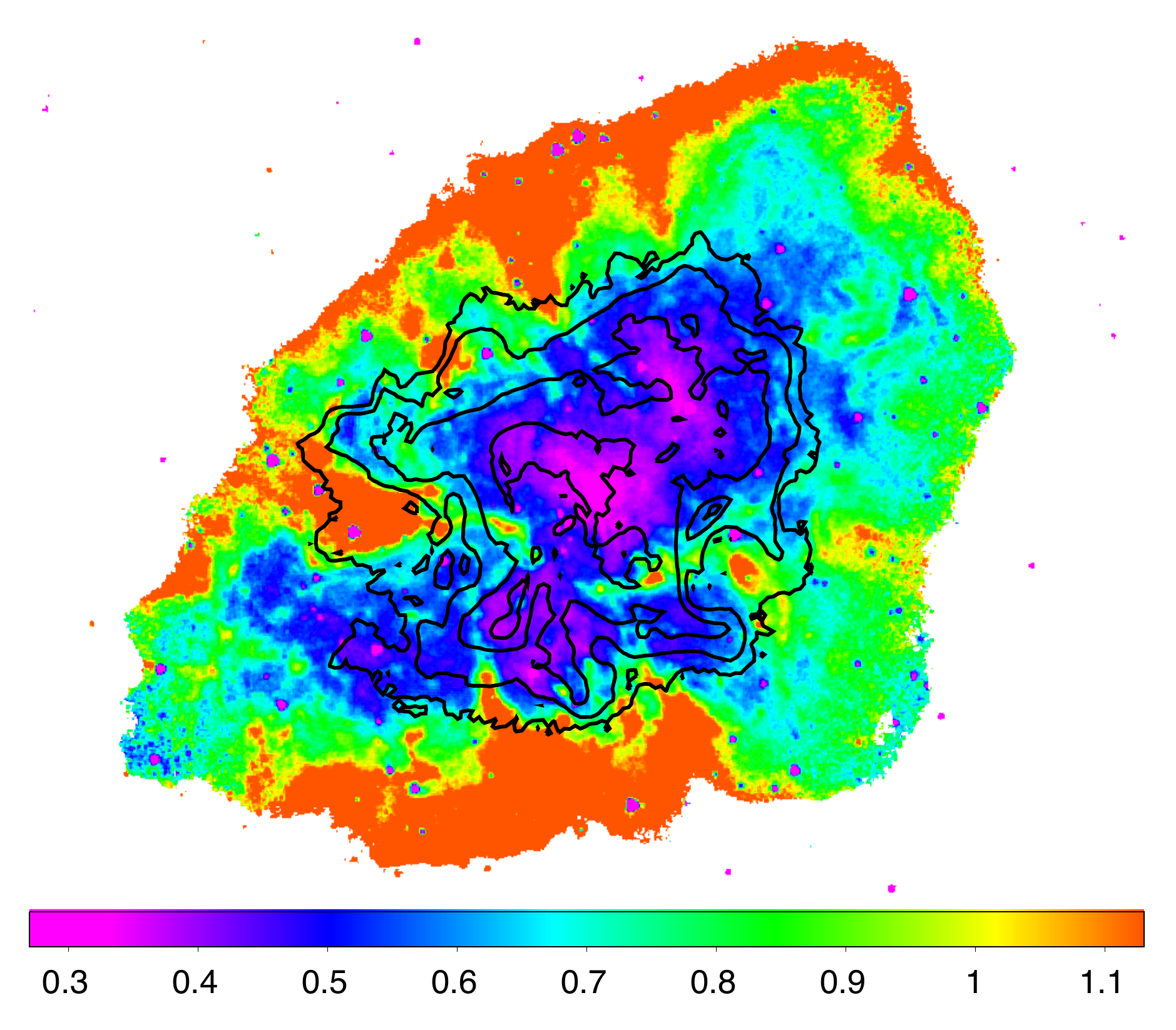}
\includegraphics[width=.49\columnwidth]{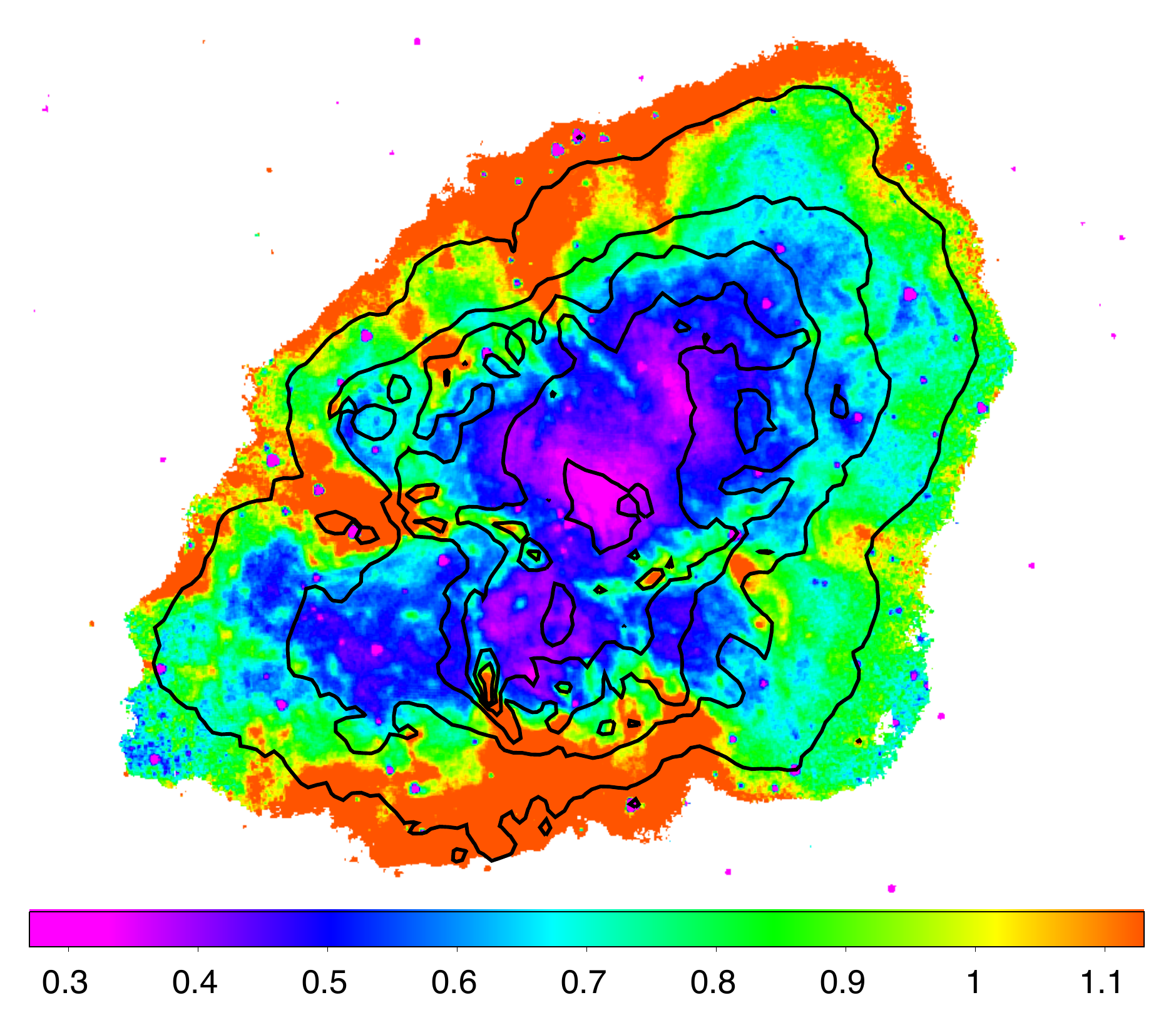}
\caption{
Left Panel: the  spectral index map from Fig.~\ref{alpha-map} (calculated as a difference
between the synchrotron-dominated $3.6 \mu$m and $4.5\mu$m images)  overlaid with the contours from the deep \textit{Chandra} X-ray image of the Crab Nebula in black \protect\citep{2006ApJ...652.1277S}. 
Right panel:  the IR spectral index map from Fig.~\ref{alpha-map}  overlaid with the contours from the intensity at $3.6\mu$m. 
}
\label{ir_xray}
\end{figure}

 Fig.~\ref{ir_xray} shows a deep \textit{Chandra} X-ray image of the Crab Nebula (contours) overlaid with  the IR spectral index map from Fig.~\ref{alpha-map} (left) in magenta. The contours appear to correlate with   the morphology of the X-ray emission, both from the inner torus and jet, as well as the fainter regions in the outer nebula.
Comparing the spectral maps in the IR with intensities in IR and in the X-rays, Fig.~\ref{ir_xray}, we find that  there is a correlation, though not a one-to-one correspondence between the IR spectral maps and X-ray intensity contours.

%

The discussion above describes a complicated picture in the IR. First, there is clear  spectral evolution towards the edges both in the optical and   in the near-IR, 
indicating that the electrons emitting at these wavelengths also have overall  cooling
times shorter than the age of the Nebula \citep{2002ASPC..271..161G,2006AJ....132.1610T}
\citep[Turbulent mixing may smooth variations of the spectral
index][]{2012ApJ...752...83T,2016MNRAS.460.4135P}. Second, there is some intensity
and spectral correlation with X-rays, but  the spectral index in the central part,
$\sim 0.3$, matches that of the radio emission (this   could indicate that we start to see the extension of the radio component). 

We interpret these somewhat contradictory facts due to the presence of two emission/acceleration components that produce nearly equal intensities in the IR, \S \S \ref{themodel}. Previously,
\cite{2002A&A...386.1044B} argued for two separate distributions,
based on significant spatial variations of the spectral index.
\cite{2002ASPC..271..161G,2006AJ....132.1610T} found   spectral
steepening towards the PWN edges, similar to that observed in the
optical band.

\section{Interpretation of spectrum: two components}
\label{Two-comp}

\subsection{Location and spectra of components}

\begin{figure}[h!]
\includegraphics[width=\columnwidth]{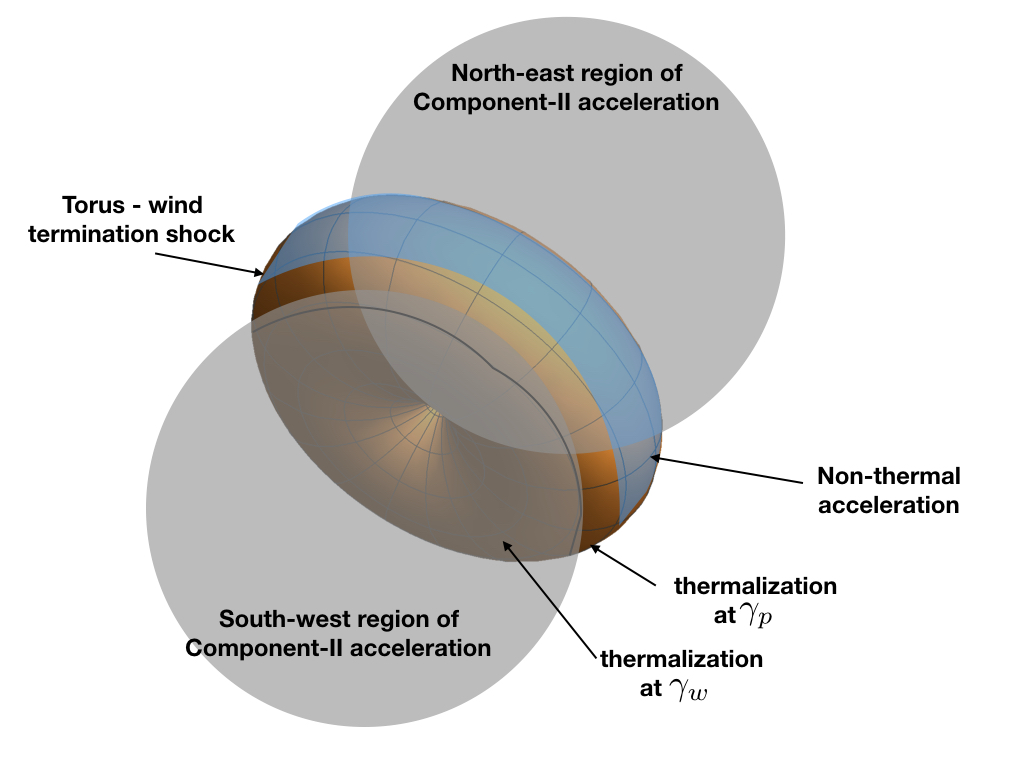} 
\caption{Qualitative picture of various acceleration regions within the Nebula. The wind termination shock is elongated in the equatorial plane. In a limited equatorial sector the shock produces non-thermal   particle spectrum via Fermi-I mechanism (blue stripe), while  at intermediate  latitudes the wind is thermalized at $\gamma_p \sim \gamma_w \sigma_w$. These sectors produce Component-I. Since the post-shock flow is mildly relativistic, one also expects NE to SW anisotropy (blue stripe).  At higher  latitudes the wind is thermalized at $\gamma_w$. In the bulk of the Nebula, closer  to the axis (still in the central part), high \Bf\ regions form (see white regions in  \protect Fig. \ref{PWN-overview}), where plasma turbulence with reconnecting current sheets accelerates Component-II  (the south-west is in the font of the torus). The gradient (from NE to SW) in the NuSTAR maps \protect\citep{2015ApJ...801...66M,2017SSRv..207..175R} possibly arise due to the fact that 
the flow  immediately post-the-equatorial shock is mildly relativistic - the blue ribbon in is NE-SW asymmetric.
}
\label{Crab-sites}
\end{figure}

\begin{figure}[h!]
\includegraphics[width=\columnwidth]{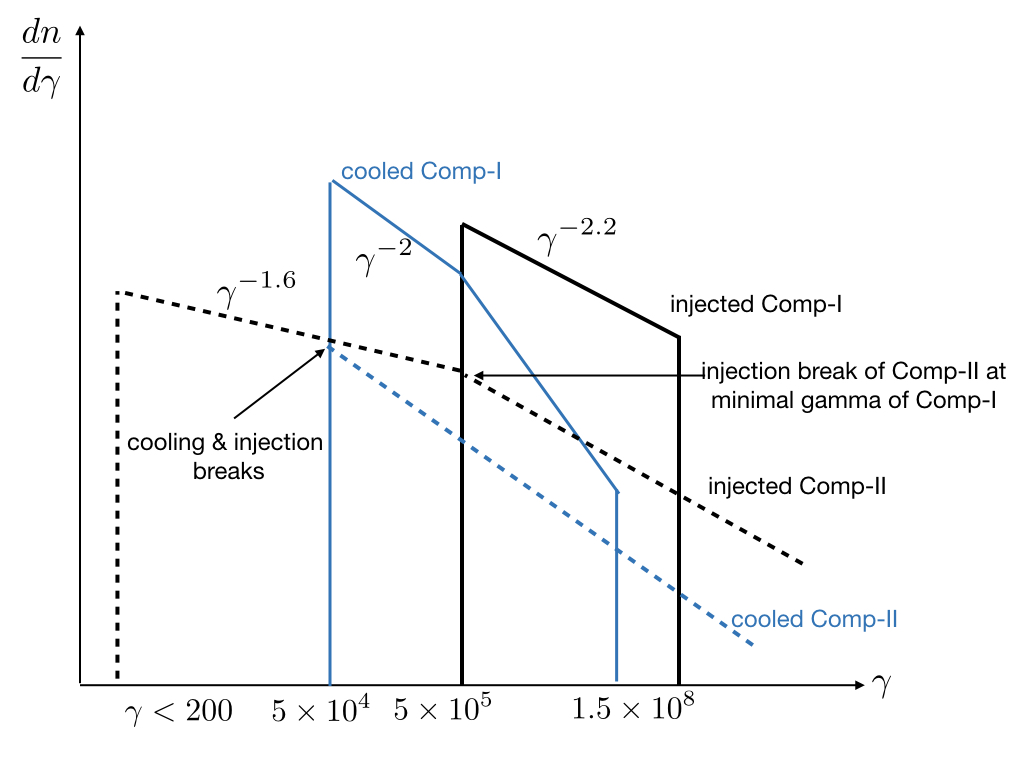}
\caption{
Qualitative picture of two particle distributions. Component-I:
solid black line - injected spectrum, solid blue line - cooled
spectrum. Component-II: dashed black line - injected spectrum,
dashed blue line - cooled spectrum. (Slope of Component-II above the break is not well constrained.) This figure illustrates the initial and eventual spectra for  each shell of particles injected into the nebula at some time in the past, not   the integral spectra of the components  accumulated over the  life of  the nebula. In addition, spacial variation of the relative intensities are expected.}
\label{Crab-particle-pictures}
\end{figure}

In  Fig. \ref{Crab-sites} we picture various acceleration regions within the Nebula, while in  Figs. \ref{Crab-particle-pictures} and Fig,
\ref{Crab-spectrum-pictures} we outline the particle distributions and the observed spectra.

  Qualitatively, there are two
components that contribute to the emission; we call them Component-I
and Component-II. Component-I is, using the terminology from Gamma Ray Bursts
\citep{1996ApJ...473..204S}, in the fast cooling regime:  the 
minimum injection energy corresponds (overall, on the time scale of the Nebula) to synchrotron emission  above the spectral break associated with
synchrotron cooling.  Component-II is in the slow
cooling regime: the injection peak is below the cooling frequency
(the higher energy tail of Component-II is still affected by
cooling).
(Here and below we use interchangeably the terms the energy of the injected particle and the corresponding energy of the synchrotron photons.)

{Our picture for the particle population in the Crab Nebula
proposes two distinct components -- one (Component-I) injected at
low latitudes  near the termination shock (corresponding to the Crab torus), and one (Component-II)
injected at higher latitudes and subsequently modified by plasma turbulence with reconnecting current sheets
 in the bulk of the Nebula. This is illustrated in Fig.
\ref{Crab-particle-pictures} where the solid (dashed) curves provide
a qualitative description of Component-I (II). Component-I is
characterized by a single power law with $dn/d\gamma \propto
\gamma^{-p_I}$ while Component-II is assumed to be a broken power
law with indices $p_{II}$ and $p_{II,above}$   below  and above  the
injection break at $\gamma_{II,br}$.  The black curves represent the
spectrum at the time of injection while the blue curves illustrate
the evolved spectra showing breaks introduced by synchrotron cooling.

Thus, an inherent break in Component-II is assumed at injection. As we argue in \S
\ref{CompoII}, we expect that particle acceleration in reconnection sheets  is sensitive
to the average energy per particle in the wind, Eq.(\ref{gammap}. The average energy
comes from two components -  bulk motion, $\gamma_{w}$ and magnetization $\sigma_{w}$. As a result, we expect that  Component-II has an injection break at the same energy as the minimal injection energy of the Component-I, $\gamma_{I, min}$, since both lie around the mean energy per particle

 Let us summarize observational evidence, though not conclusive,  for the
Component-II above the IR:
 \begin{itemize}
 \item  Indications of a spectral break around $130$  keV in the overall spectrum of
the Crab, \S  \ref{COMPTEL}.
  \item  Indications of spectral bump (COMPTEL bump) in the $\geq$ MeV region, \S  \ref{COMPTEL}.
\item The NuStar spectral map above 10~ keV shows that the hardest spectra are not
correlated with the $\leq 10$  keV X-ray torus, \S \ref{Xrays}. (NuStar sensitivity
above 10~ keV  extends to $\sim 70 $  keV, so this statement applies to that energy.)
\item In IR the spectral index of the central part of the Nebula ($\sim 0.3$) matches
that of the radio photons, Fig. \ref {alpha-map}.
\end{itemize}

\begin{figure}[h!]
\includegraphics[width=.95\columnwidth]{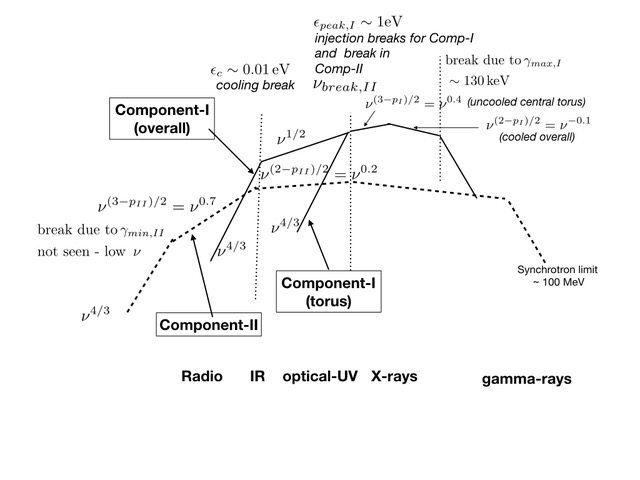} 
\caption{
Qualitative picture of two synchrotron components in $\nu F_\nu$. Component-I
(solid line) is in the fast cooling regime, with injection energy above
the cooling break, while Component-II (dashed line) is in the slow
cooling regime.   In projection Component-II  partially overlaps with the torus. The spectral breaks are: (i) cooling breaks in
both components ($\nu_c$, energy approximately $ 0.01 $ eV). [The
transition between dashed ($\nu^{0.7}$, Component-II dominated)  to
solid ($\nu^{0.5}$, Component-I dominated) is not as sharp as is
drawn, since the cooling frequency is a very sharp function of the local
B-field; plus there is large contribution from dust.]  (ii) UV peak
- this is $\gamma_{min, I}$ of Component-I (the wind Lorentz
factor); (iii) possible $130$  keV break due to $\gamma_{max, I}$;
(iv) hypothetical break in Component-II somewhere in the UV-soft
X-rays -- this is required for Component-II not to overshoot
Component-I; (v) Synchrotron emission limit, $\sim 100$ MeV.
 It is expected that variations of the \Bf\   and of the dynamical times (\eg, in the inner and outer parts of the Nebula)  lead to further complications of the picture.
}
\label{Crab-spectrum-pictures}
\end{figure}

The associated synchrotron emission from each particle component
is illustrated in Fig. \ref{Crab-spectrum-pictures} where we
identify spectral indices and breaks associated with the evolution
of the particle spectra, and constrained by observations of the
broadband spectrum of the Crab Nebula. The associated spectral
breaks (numbered as in \S \ref{breaks}) are (see also Table \ref{table}; the numbers given are approximate estimates and may not be unique): 
\begin{enumerate}
\setcounter{enumi}{-1}

\item Injection break for Component-II, $\gamma_{min, II}$ (not observed - corresponding  synchrotron frequency below $\sim $ few tens of MHz)

\item Radio: the required particle distribution of synchrotron
emitting leptons (uncooled Component-II) is $p_{II} =1+2 \alpha_r = 1.6$ (by
$p$ we mean the injected spectrum).

\item Cooling break in mid-IR. Spectral steepening towards the edges in the optical \cite{1993A&A...270..370V} and near-IR \cite{2002ASPC..271..161G,2006AJ....132.1610T}
indicated that optical and near IR electrons are in a fast cooling regime - the cooling frequency is in the mid-IR.
 Particles enter the fast cooling regime, showing steepening towards the edges. The intensity of Component-I sharply drops below the cooling break, where $\nu F_\nu \propto \nu^{4/3}$; above the cooling break Component-I dominates over Component-II. 

\item In the optical $\alpha_o \approx 0.5$ -- these are the electrons from Component-I that were initially emitting in the UV and X-rays, in a fast cooling regime, but cooled down to emit in the optical. In the optical there is a clear steepening of spectral index towards the edges \cite{1993A&A...270..370V}. 

\item UV peak - $\gamma_{min, I}$ - the peak emission of Component-I. This is the peak energy per particle in the equatorial part of the wind, $\gamma_{min, I} = \gamma_p$.

\item We interpret the break at $\sim 130$  keV (if it is real) as $\gamma_{max, I}\sim 1.5 \times 10^8$, Eq. (\ref{gammamaxI}), of Component-I.  If the claim of the break is not supported by observations, there is no limit on $\gamma_{max, I}$ .

\item MeV bump - possible reappearance of Component-II. (Thus, above few hundred  keV, we argue, we start to see the extension of the radio component.) If the MeV bump is not supported by observations, Component-II has a substantial spectral break between IR and X-rays.

\item Synchrotron burn-off limit \citep{1996ApJ...457..253D,2010MNRAS.405.1809L}.

\item IC component \citep{1996MNRAS.278..525A,1996ApJ...457..253D}.
\end{enumerate}

\begin{table}
\caption{Properties of the two components.}
\begin{tabular}{l|ll}
\hline
Quantity & \Lf\ & photon energy \\
\hline
$\gamma_{min, I}$ - minimal injection for Comp.-I &$ \sim 5 \times 10^5$& $\sim $eV\\
$\gamma_{max, I }$ - maximal injection for Comp.-I &$ \sim 1.8 \times 10^8$&$ \sim 100$ keV\\
$\gamma_{c} $- cooling energy (both Comp.-I and  Comp.-II) &$ \sim 5 \times 10^4$&$ \sim 0.01$ eV\\
$\gamma_{min, II} $ - minimal injection for Comp.-II &$\leq 200 $&$ \leq 100 $MHz\\
$ \gamma_{II,br}  = \gamma_{I, min} $ - break in  injection  spectrum for Comp.-II &$\sim 5 \times 10^5$&$ \sim$ eV\\
\hline
\end{tabular}
\label{table}
\end{table}

Thus, our interpretation of the spectrum differs from
\cite{1984ApJ...283..710K} in one particularly important way:
\cite{1984ApJ...283..710K} (see also \cite{2011MNRAS.410..381B})
interpreted the UV break as a cooling break, and the IR break as
an injection break. We argue it's the reverse - both IR and optical spectral maps show clear evolution towards the edges, implying that IR and optical electrons have cooling time scale shorter than the life time of the Nebula.

  \subsection{Component-I}
  
   \subsubsection{Wind \Lf\ and magnetization}
  The Inner Knot in the Crab Nebula -- a small bright spot close to the pulsar
\citep{2015ApJ...811...24R}  is interpreted as a surface of the relativistic pulsar wind  termination shock \citep{2016MNRAS.456..286L,2015MNRAS.454.2754Y}. Thus, its properties can be used to probe the properties of the pulsar wind.  The Inner Knot is located at intermediate latitudes, $\sim 45^\circ$ off the spin axis. Importantly, the spectrum of the 
   Inner Knot is thermal, 
 with a peak
frequency $\nu_{knot} \approx$  few times $ 10^{14}$ Hz  \cite[][ though this result
has not been confirmed by others yet]{2017SSRv..207..137P}.  The thermal nature of the
knot is, first, consistent with the model of \cite{2011ApJ...741...39S} (see also \S \ref{PICI})
that predicts that the termination shock in the striped wind does
not accelerate particles outside of the $\sim \pm 10^\circ$ sector
from the equatorial plane. Second, using the model of the post-shock
flow developed by \cite{2016MNRAS.456..286L}, one can estimate the post-shock energy per particle $\gamma_p$.
 Using the peak frequency of the
knot emission, and estimates of the post-shock \Bf\ and \Lf\, one finds
  \be
  \gamma_p = \left( \frac{ 2 \pi m_e c \nu_{knot} }{ e B}\right)^{1/2} \approx 5 \times 10^5 b_0^{-1}
  \label{gammaw}
  \ee
  (Recall that the average energy per particle  in  the wind  consists of the wind \Lf\
times the magnetization, Eq. (\ref{gammap})).
Below we will use the estimate (\ref{gammaw}) as the minimal $\gamma_{I, min} $ for the 
Component-I.

The peak energy of Component-I is then
\be
\epsilon_{p,I} \approx \hbar \gamma_p^2 \om_B = 1.5 b_0 \, {\rm eV}
\ee
{where $\om_B= e B/(m_e c)$ is the non-relativistic cyclotron frequency}. This matches the observed peak 
in the Crab spectrum.
  
The overall  cooling break is in the IR:
\ba &&
\epsilon_c = \frac{ m_e^5 c^9 \hbar}{e^7 B^3 \tau^2}= 10^{-2} b_0 ^{-3}\, {\rm eV},
\label{epsilonc1}
\\ &&
\gamma_c = \frac{m_e^3 c^5}{e^4 B^2 \tau}= 5 \times 10^4 b_0^{-2} ,
\label{epsilonc}
\ea
where $\tau$ is the age of the Nebula. 

Note the sharp dependence of $\epsilon_c$ and $ \gamma_c $ on the local
\Bf. In a fluctuating field this will produce large variations
in the cooling energy; variations of the \Bf\ by a factor of $\sim
2$ would produce variations in the cooling break of an order of
magnitude -- the cooling frequency $\nu_c$ is expected to vary by
at least an order of magnitude between different regions in the
Nebula.

Thus, the cooling energy calculated with the age of the Nebula is below the peak energy, $\epsilon_c < \epsilon_{p,I}$. According to  the terminology
of Gamma-Ray Bursts this regime ($\epsilon_c \leq \epsilon_p$) is
called fast cooling \citep{1996ApJ...473..204S}.
 In the fast cooling
regime,
\be
\nu F_\nu \propto
\left\{
\begin{array}{cc}
\epsilon^{4/3} & \epsilon < \epsilon_c
\\
\epsilon^{1/2} &\epsilon_c <\epsilon < \epsilon_{p,I}
\\
\epsilon^{(2-p_I)/2} \propto \epsilon^{-0.1}&\epsilon_{p,I} <\epsilon 
\end{array}
\right.
\label{pI}
\ee

On the other hand, the inner parts of the Nebula -- especially the torus -- have
dynamical times of few months, and are thus in the slow cooling regime. Thus we expect for the torus
\be
\nu F_\nu \propto
\left\{
\begin{array}{cc}
\epsilon^{4/3} & \epsilon <  \epsilon_{p,I}
\\
\epsilon^{(3-p_I)/2} \propto \epsilon^{0.4}&\epsilon_{p,I} <\epsilon 
\end{array}
\right.
\label{pI1}
\ee
plus a possible exponential break corresponding to $\gamma_{I,max}$,
and synchrotron limit at $\sim 100$ MeV or less; in (\ref{pI}-\ref{pI1}) the injection
power-law is $ p_I \approx 2.2$.

If the $\epsilon_b \sim 130 $  keV break \citep[][and \S \protect
\ref{COMPTEL}]{2010A&A...523A...2M} is real, this can be an indication
of the maximal acceleration limit of Component-I, $\gamma_{max,
I}$:
\be
\gamma_{max, I}= \left( \frac{ m_e c\epsilon_b}{e \hbar B} \right)^{1/2}= 1.5 \times 10^8 b_0^{-1/2}
\label{gammamaxI}
\ee



\subsubsection{The torus: Component-I in slow cooling regime in the central region}
\label{torusoptical}

In the overall spectrum of the Nebula, Fig \ref{Crab-spectrum-pictures-1}, the near-IR flux of few $\times 10^{-8}$ erg cm$^{-2}$ s $^{-1}$ is only mildly lower than in X-rays,
$\sim 10^{-8}$ erg cm$^{-2}$ s $^{-1}$. Yet the torus is barely seen in IR, see \S \ref{internal}. Thus the torus emission should decrease sharply below the peak. This can be achieved only if the UV peak is the injection break in the slowly cooling regime \footnote{Overall, on the lifetime of the Nebula, Component-I is in the fast cooling regime, while the much smaller torus is in the slow cooling regime.} - then the slope of $\nu F_\nu$ changes from $4/3$ below the peak to $(3-p_{I})/2$ and then to $(2-p_{I})/2$) above the peak.

Consider the immediate post-termination shock region (to be called the ``torus'') as a
separate entity within a Nebula. Accelerated particles are injected into it
(Component-I), they emit, and are advected out by the flow. Thus, the number of emitting
particles within the torus region is constant in time (for approximately constant spin-down luminosity). Let's estimate the expected ratio of intensities at two different wavelengths. 

As a comparison point, we take X-rays, where the emitting particles are clearly above
the injection break and are fast cooling. For frequencies below X-rays, but above the
injection break (still in fast cooling), $F_\nu \propto \nu^{-p/2}$. Then comes the UV break at $\nu_{break}$. If it is due to cooling then below it
\be
\frac{\left(\nu F_{\nu}\right)_{IR}}{\left(\nu F_{\nu}\right)_X}= 
\left(\frac{\nu}{\nu_{break}}\right)^{(3-p)/2}\left(\frac{\nu_{break}}{\nu_X}\right)^{1-p/2}
\approx 0.65
\label{FF}
\ee
for $h \nu= 0.1$ eV, $h \nu_{break}= 1.5$ eV, $h\nu_X= 10^3$ eV and $p=2.2$.

Thus, in this case we would expect that the torus is as bright in optical and IR  as in X-rays (in terms of surface brightness).

Alternatively, the UV break at $\nu_{break}$ can be due to injection. Then there are two cases:
(i) If the residence time is sufficiently short, so that even particles with $\gamma_{min, I }$ do not experience cooling, then
\be
\frac{\left(\nu F_{\nu}\right)_{IR}}{\left(\nu F_{\nu}\right)_X}= 
\left(\frac{\nu}{\nu_{break}}\right)^{4/3}\left(\frac{\nu_{break}}{\nu_X}\right)^{1-p/2}
\approx 0.05
\label{FF1}
\ee
Thus, the expected torus brightness in this case is more than an order of magnitude smaller.

(ii) If particles with $\gamma_{min}$ do cool to below the observed frequency, then
\be
\frac{\left(\nu F_{\nu}\right)_{IR}}{\left(\nu F_{\nu}\right)_X}= 
\left(\frac{\nu}{\nu_{break}}\right)^{1/2}\left(\frac{\nu_{break}}{\nu_X}\right)^{1-p/2}
\approx 0.5,
\label{FF2}
\ee
just somewhat smaller than (\ref{FF}).

Which of the estimates (\ref{FF1}) and (\ref{FF2}) is closer to reality? The overall
estimate of the cooling break (\ref{epsilonc}) is based on the age of the Nebula. But the
inner regions of the Nebula have dynamical times much smaller than the age of the Nebula. As a result, the cooling break for the newly injected particles in the inner parts of the Nebula, in the torus, is at higher energies. Thus, we expect that particles with $\gamma_{min, I }$ will not cool to below ($0.1$ eV). Hence, estimate (\ref{FF1}) is applicable.

These simple theoretical estimates demonstrate that intensity of the torus can
decrease precipitously if we identify the UV spectral break with the injection break of
Component-I. This then explains the sharp decrease of the torus emissivity from the UV
towards the IR.

 In principle, this {\it does not mean} that radio electrons are not accelerated at the termination shock. If a ratio of radio and X-ray emitting electrons within the same volume is $n_R/n_X$, then in the same \Bf\
the ratio of resulting fluxes is
\be
\frac{F_R}{F_X}= \frac{n_R}{n_X} \frac{\nu_R}{\nu_X}
\ee
Since $\nu_R/{\nu_X}\sim 10^{-9}$ - very little emission is expected from the newly injected radio particles: radio-emitting leptons are highly inefficient radiators; the overall radio emission is dominated by the accumulated long-living leptons.

In conclusion: radio electrons may be accelerated at the termination shock, but there is no observational requirement that they {\it are} accelerated at the termination shock; the shock may not accelerate any radio electrons.

 \subsubsection{Component-I: energetics}
 
 How do Component-I and Component-II compare energetically? 
Overall, most of the radiative power comes out in the UV, due to Component-I.
 In our model the peak UV power  of Component-I comes from two related, but
somewhat separate components: (i) power-law acceleration by Fermi-I process in the
equatorial part, that extends from UV to X-rays -- this is energetically the dominant
component;  (ii) thermalization at the high latitudes that produce a broad peak in  the UV. 
  The power of Component-I is related to the current spin-down  power of the pulsar (since Component-I is in the fast cooling regime, most of the energy given to particles at the termination shock is radiated away).

   The radiative efficiency of Component-I in case of the Crab is a few percent. For the   wind power $L_w \propto \sin^2 \theta$ ($\theta$ is the polar angle), the amount of wind energy that comes within $\pm 10^\circ$ of the equatorial plane is $\sim 25\%$.  Consistent with this estimate, the peak luminosity of Component-I is $ L_{p,I} \sim 10^{37}$ erg s$^{-1}$, see Fig. \ref{Crab-spectrum-pictures-1}, safely below the spin-down power of the Crab pulsar.

 Component-I peaks in the UV at frequencies corresponding to $\gamma_{min, I}$; it cools down to optical-IR over the lifetime of the Nebula. Thus, 
 Component-I is in the fast cooling regime: most of the energy given to the particles during acceleration process is radiated away. 
To estimate the radiative efficiency of Component-I, we note that the cooling time for the $\sim 1$ eV injection peak is
\be
\tau_{c,peak} = \frac{m_e^{5/2} c^{9/2} \sqrt{\hbar}}{e^{7/2} B^{3/2} \sqrt{\epsilon_p} }\approx 100\, {\rm yrs}\, b_0^{-3/2}
\label{taucpeak}
 \ee
 This estimate matches really well the fact that there is a strong spectral evolution in the optical \citep{1993A&A...270..370V}.
 
The estimate (\ref{taucpeak}) explains the radiative efficiency of the Inner Knot: \cite{2016MNRAS.456..286L} found that its radiative efficiency is $\sim 10^{-3}$. 
 Since its spectrum peaks approximately at $\sim $ eV \citep{2017SSRv..207..137P},
similar to the peak of the overall spectrum in the Crab, this factor of $\sim
10^{-3}$ is naturally explained as a ratio of the dynamic time of the Knot, $\sim$
few month \citep{2015ApJ...811...24R}, to the cooling time scale (\ref{taucpeak}).
After a few months the directed post-shock flow loses directionality and becomes
turbulent, so that the optical-UV emission is no longer contained within a relatively bright Inner Knot \cite[see][for discussion of how extended is the post shock emission in the Inner Knot]{2016MNRAS.456..286L}.
  Importantly, the whole flow contributes to the UV peak of Component-I, not only the equatorial belt where acceleration of X-ray emitting electrons occurs. 
 
In conclusion, the UV peak has cooling time scale moderately small compared to the
age of the Nebula -- thus any small feature like the Inner Knot is not very bright.
Most of the emission will be distributed. On the other hand, in X-rays the cooling
time scales are similar to the dynamic time scales, only a few months, producing sharp features (the X-ray torus). 

Overall, the peak intensity in $\nu F_\nu$ of a few percent seems like a good consistent estimate of the radiative energetics of Component-I.

  \subsection{Component-II}
  \label{multiplicity}
  
  \subsubsection{The spectrum}

The emission corresponding to the minimal injection energy of Component-II is very low, below $\leq 100$
MHz, so that $\gamma_{p, II} \leq 200$ \citep[the radio power-law
spectrum extends to at least 30MHz][]{1971IAUS...46...22B}.  The
cooling energy is the same as for Component-I, Eq. (\ref{epsilonc}).  Thus, Component-II
is in the slowly cooling regime ($\epsilon_c \geq \epsilon_p$).

In addition,  Component-II has an intrinsic injection break at the minimal energy of Component-I.
 In
this case
  \be
\nu F_\nu \propto
\left\{
\begin{array}{cc}
\epsilon^{4/3} & \epsilon < \epsilon_{p,II}
\\
\epsilon^{(3-p_{II})/2 }= \epsilon^{0.7} &\epsilon_{p,II} <\epsilon < \epsilon_c
\\ 
\epsilon^{(2-p_{II})/2 }= \epsilon^{0.2}& \epsilon_c <\epsilon <\epsilon_{p,I}
\\ 
\epsilon^{(2-p_{II,above})/2 }= \mbox{(steeper than)} \epsilon^{-0}&\epsilon_{p,I}<\epsilon 
\\
e^{-\epsilon/\epsilon_s} & \epsilon_s < \epsilon, \,  \epsilon_s \sim 100 MeV
\end{array}
\right.
\label{pII}
\ee
with $p_{II}=1.6$ and $p_{II,above}> 2$

\subsubsection{Component-II: energetics}

The luminosity of Component-II  at the cooling break is $\nu F_\nu \sim 10^{-8} $ erg cm$^{2}$ s$^{-1}$, see Fig. \ref{Crab-spectrum-pictures-1} (it grows slowly above the cooling break). So, 
the present total luminosity of Component-II is $L_{p,II} \sim 5 \times 10^{36}$ erg
s$^{-1}$, well below the spin-down luminosity (this estimate may increase due to a
possible rise of Component-II towards the optical). The total energetics of Component-II
for a distance of 2~kpc and age of 1000 yrs is $\sim 10^{48}$ ergs. This is  a considerable fraction (few percent at least) of the total
energy of the MHD flow \citep{1984ApJ...283..694K}. Since, in the IR, electrons are in a marginally cooled state, this is also the estimate of the total 
energetics in the particles contributing to Component-II. It is safely below the power that the pulsar put into the Nebula during its life time, regardless of the initial spin.

 In other words, assume that Component-II continues  without a break up to the
injection energy of Component-I, $1$ eV, but  only the $0.01-1$~ keV part is in the fast cooling regime. In a \Bf\ $b_0$ the corresponding {\Lf}s are 
  $\gamma_p \approx 5 \times 10^5$,  $\gamma_c \approx 5 \times 10^4$. Neglecting for simplicity the energy of Component-II above the break, the relative  amount of energy deposited into the fast cooling part of Component-II  is then
  $1- (\gamma_c/\gamma_p)^{2-p_{II}} \approx 0.6$ for $p_{II}=1.6$. Thus, approximately half of the energy transferred from the \Bf\ to the particles in the reconnection events (Component-II) is given to the slowly cooling particles. 
  
Thus, Component-II dissipates  a considerable fraction of the Poynting luminosity of the wind (and most of the magnetic flux). A large fraction of the injected energy remains in the form of the turbulent \Bf, plus approximately half of the dissipated energy is given to the slowly cooling particles. We thus conclude that the overall energetic requirement on Component-II are consistent with observations.

The fact that the energetics of Component-II is a non-negligible fraction of the total
energy put into the Nebula by the pulsar is expected on theoretical grounds -- in order to resolve the  sigma-paradox a considerable fraction of the spin-down luminosity should be dissipated \cite{2006NJPh....8..119L,2013MNRAS.431L..48P}.
(As a cautionary note, what is needed to resolve the sigma-paradox is the destruction of the large scale
 magnetic flux, not magnetic energy \cite{2006NJPh....8..119L}.)

Thus, the energetics of Component-II is very reasonable -- a large fraction of the energy
that the pulsar puts into the Nebula in a form of mostly Poynting flux is dissipated in reconnection events within the turbulent medium. On the other hand, the problem of the number of radio electrons remains, provided they are supplied by the wind \citep{atoyan_99,2013MNRAS.428.2459K}.
 For the number distribution $dn/d\gamma = C \gamma^{-p_{II}}$, and assuming that the distribution extends to at least the cooling break, the total number of electrons within Component-II can be estimated as
 \be
 N_{II} \approx \frac{L_{p,II} \tau}{m_e c^2} \gamma_c ^{-2+ p_{II}} \gamma_{min,II} ^{1- p_{II}} = 
 \frac{L_{p,II} \tau}{m_e c^2} \gamma_c^{-0.4} \gamma_{min,II}^{-0.6}= 4\times 10^{51} \, b_0^{0.8} \gamma_{min,II}^{-0.6} \, .
 \label{NII}
 \ee
 Comparing with the injection rate
\be
\dot{N} = \lambda \frac{\Omega B_{NS}}{2 \pi e c} \pi R_{NS}^2 \frac{\Omega R_{NS}}{c} c
\ee
the needed multiplicity is
\be
\lambda = 2 \times 10^7 b_0^{0.8} \gamma_{min}^{-0.6} \, .
\ee
The allowed $\gamma_{min} \geq 100$, which would still require $\lambda \geq 10^6$. This is $\sim$ two orders of magnitude higher than the models of pulsar magnetospheres predict \citep[\eg][]{AronsScharlemann,2010MNRAS.408.2092T}.

\subsubsection{Can Component-II extend to GeV energies?}

Synchrotron spectra of accelerated particles may generally have two  cutoffs - the synchrotron limit \citep{1996ApJ...457..253D,2010MNRAS.405.1809L} and possibly a potential limit within the acceleration region.
The synchrotron limit (independent of \Bf) is given by Eq. (\ref{emax}).

The total electric potential drop with the Nebula corresponds to the maximal \Lf\ 
\be
\gamma_{tot} \sim \frac{e \sqrt{L_{sd}}}{m_e c^{5/2}} = 7\times 10^{10}
\label{gammamax}
\ee
where $L_{sd}= 4 \times 10^{38} $ erg s$^{-1}$ is the spin-down power. It is hard/nearly impossible for particles, especially leptons, to gain the full potential
\citep[][discussed how cosmic rays with large Larmor radius can achieve the potential drop]{1992MNRAS.257..493B}.

If a particle gains a fraction $\eta_p$ of the maximal potential (\ref{gammamax}), then
in a given \Bf\ particles  will produce synchrotron photons with energy
\be
\epsilon_{tot} \approx \hbar \eta_p^2 \gamma_{max}^2 \om_B = \frac{e^3 \hbar B L_{sd}}{m_e^3 c^6}= 20 \, {\rm GeV} b_0\, \eta_p^2
\ee
Thus, to reach the synchrotron limit (\ref{emax}) a particle needs to cross $\sim 0.05$ of the total potential \citep{2018JPlPh..84b6301L}.

The example of Crab flares demonstrates that under certain conditions particles do achieve the synchrotron limit (\ref{emax}). Thus, there are acceleration sites that can, at least occasionally, gain large overall potentials, with $\eta_p \sim 0.1$. This implies that there is no intrinsic limit on $\eta_p$ (that is, there is no requirement from observations to have an intrinsic limit on the possible accelerating potential - there is no limit on $\eta_p$, besides $\eta_p< 1$). 

One can then imagine a (power-law) distribution of reconnection sites with different total electric potential drops $\eta_p<1$, with no special value for $\eta_p$. The overall spectrum is then a combination of intrinsic $p$ (that may also depend on the distribution of $\sigma_w$), distribution of  $\gamma_{max}$, and how many particles are accelerated in a sheet with given $\gamma_{max}$.


\subsubsection{ Dominance of Component-II in radio}

Scaling (\ref{pI}) and (\ref{pII}) naturally explain why Component-II dominates in
radio -- below the cooling frequency (globally in the IR, (\ref{epsilonc}), but locally at higher frequencies) Component-I sharply drops off  (as $\epsilon^{4/3}$ in $\nu F_\nu$) and becomes subdominant to Component-II. 
 Component-I dominates in the optical, UV and X-rays (Component-II experiences a cooling break, while Component-I increases toward the injection break).
 Thus, in our model it's a natural consequence that the X-ray bright Component-I
becomes subdominant below the cooling frequency (locally in the ``torus'' around $0.1$ eV, and globally around $0.01$ eV).

\subsubsection{Explanation of the possible MeV bump/requirement of the spectral break in Component-II }
\label{bump1}

Above the injection break Component-I is slowly decreasing with $\nu F_\nu \propto \epsilon^{1-p_{I}/2} \approx \epsilon ^{-0.1}$, while below the injection break Component-I 
has  globally $\nu F_\nu \propto \epsilon^{1/2}$. In the torus   $\nu F_\nu \propto \epsilon^{4/3}$ (in the torus particles did not have time to cool to form  $ \epsilon^{1/2}$ tail).  The   Component-II
above the cooling break  is slowly increasing with  $\nu F_\nu \propto  \epsilon^{1-p_{II}/2} \approx \epsilon ^{0.2}$. Thus Component-II may start to dominate over Component-I. Parametrizing spectral fluxes as:\\
Component-I:
\be
F_I = F_{I,0} \left(\frac{ \epsilon_p}{\epsilon_c}\right)^{-1/2} \left(\frac{ \epsilon}{\epsilon_p}\right)^{-\alpha_{I}}
\ee
where $ F_{I,0}$ is the spectral flux of Component-I at the cooling energy $\epsilon_c$; above $\epsilon_c$ the spectral flux first increases $ \propto \epsilon ^{1/2}$, and then decays as $ \epsilon^{-\alpha_{I}}$. \\
Component-II:
\be
F_{II} = F_{II,0} \left(\frac{ \epsilon}{\epsilon_p}\right)^{-\alpha_{II} -1/2}
\ee
where $ F_{II,0}$ is the spectral flux of Component-II at the cooling energy, $\alpha_{II}$ is the uncooled spectral index, and factor $1/2$ in the spectral exponent accounts for the spectral break.

The two components become equal at
\ba &&
\epsilon \sim \left(\frac{ F_{I,0} }{F_{II,0} }\right)^{2/(2 (\alpha_{I} +\alpha_{II} )-1)} \epsilon_p^{(1+ 2 \alpha_{I} )/(2 (\alpha_{I} +\alpha_{II} )-1)}
\epsilon_c^{- 2(1-  \alpha_{I} )/(2 (\alpha_{I} +\alpha_{II} )-1)}
\nn &&
\approx \left(\frac{ F_{I,0} }{F_{II,0} }\right)^{3.3} \frac{\epsilon_p^{2}}{\epsilon_c}
= 200\,\rm{eV} \left(\frac{ F_{I,0} }{F_{II,0} }\right)^{3.3} b_0^5
\label{ratioo}
\ea

This is clearly not seen, yet the ratio of the intensities in the X-ray band is a
sensitive function of both the overall normalization and, especially, of the local \Bf, $\propto b_0^5$.
The ratio (\ref{ratioo}) can easily extend beyond hundreds of  keV for ${ F_{I,0} }/F_{II,0} \sim $ few and $b_0 \sim $ few. It is tempting to associate the possible COMPTEL MeV bump, \S \ref{COMPTEL} with reemergence of Component-II.

Alternatively, in order not to overshoot Component-I in the X-rays Component-II should experience a mild spectral break somewhere between IR and hard X-rays. As we argue in \S \ref{CompoII} one might expect injection the break in Component-II corresponding to the average energy per particle, $\gamma_p$, which is the $\gamma_{I< min}$
 \subsubsection{Component-II: acceleration sites}
 
 In our model Component-II is accelerated in the bulk of the Nebula, but still in the
central parts, see Fig. \ref{PWN-overview}. There are several arguments in favor of
centrally concentrated acceleration of Component-II, as opposed to evenly distributed
over the Nebula: (i)  the spectrum of Component-II requires  regions with high
magnetization, $\sigma \gg 1$ - see \S \ref{CompoII}. Such regions do exist in the
central parts of the Nebula, see Fig.  \ref{PWN-overview}; (ii)  spectral evolution in
the IR towards the edges of the Nebula, Fig. \ref{alpha-map}, requires that particles are
centrally accelerated, and that turbulence is not effective in mixing particles with
different ages.

Even though radio emitting electrons (Component-II) are accelerated now in intermittent
reconnection events, we do not expect any kind of scintillation over the radio nebula -  radio electrons are very inefficient radiators, so an addition of freshly accelerated leptons to large pool of already present ones will not produce much of an enhancement, see \S \ref{torusoptical} and \S \ref{whence}.

\begin{figure}[htbp]
\begin{center}
\includegraphics[width=0.99\textwidth]{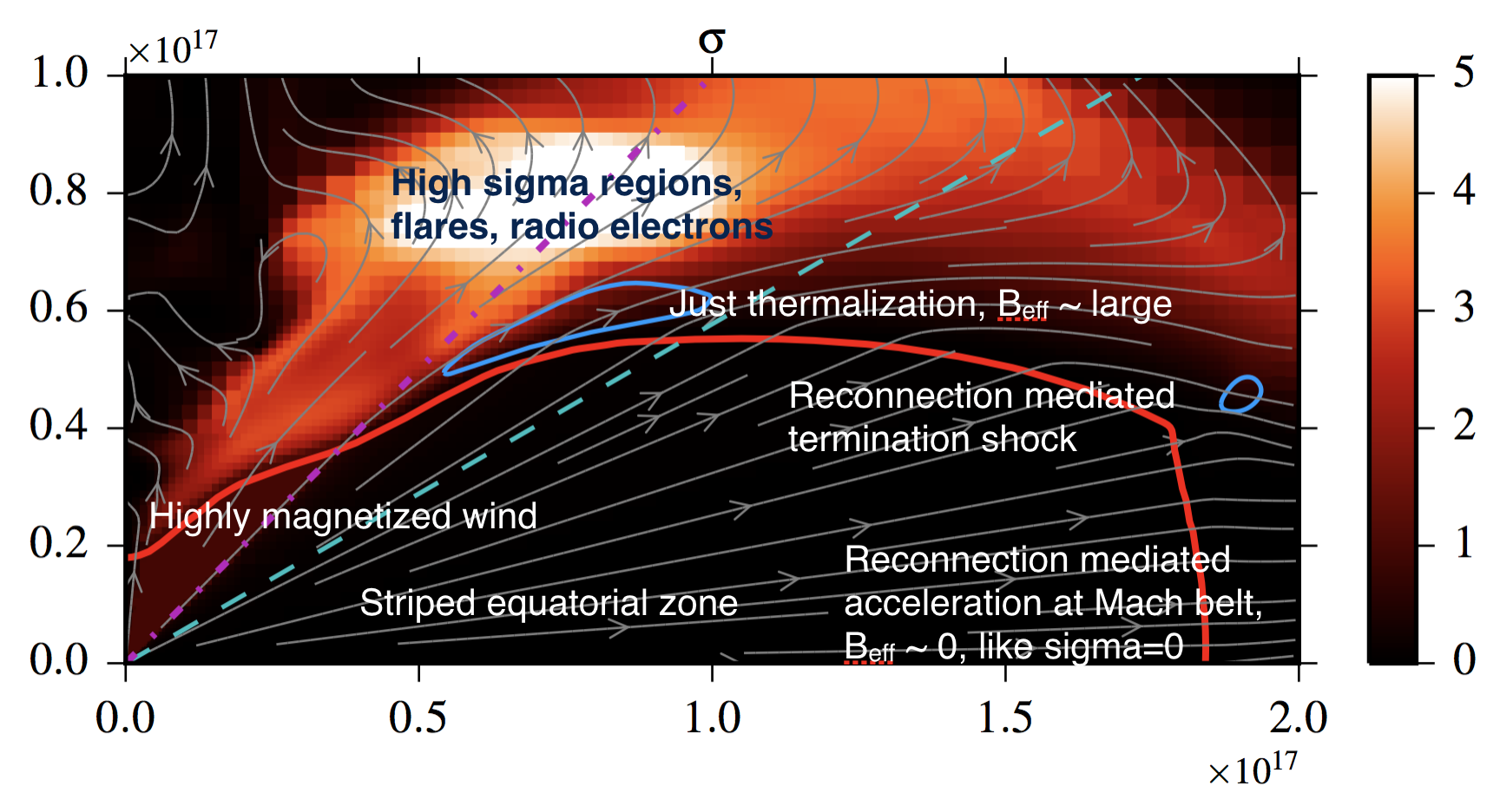}
\caption{Magnetization $\sigma_w$ and instantaneous streamlines near the termination shock in RMHD simulations of the Crab Nebula \citep{2014MNRAS.438..278P}. The dot-dashed straight line shows the separation of the polar and striped zones of the pulsar wind. The dashed straight line is the line of sight. The solid red line shows the termination shock and the solid blue loop between the dashed and dot-dashed lines shows the region of Doppler-beamed emission associated with the inner knot of the Nebula. The polar beam corresponds to the streamlines originating from the inner part of the termination shock located to the left of the intersection with the dot-dashed line. We added notations describing various properties of the wind/acceleration regions.}
\label{PWN-overview}
\end{center}
\end{figure}

\section{Origin of Component-I and Component-II}

Next we discuss a physical origin for Component-I and Component-II. Briefly, both components are supplied by the pulsar: the equatorial section of the termination shock accelerates Component-I, higher latitudes of the shock just thermalize the flow, while Component-II is given by particles  that are accelerated in the turbulence that self-consistently generates reconnecting current sheets in the bulk of the Nebula.

\subsection{Component-I: the wind termination shock}

\subsubsection{Component-I:  insights from MHD simulations}

The overall properties of the Crab X-ray emission are well reproduced by numerical models
that assume low-sigma, equatorially-dominated flow \citep{2004MNRAS.349..779K}. 
In addition, modeling of the Inner Knot \citep{2016MNRAS.456..286L,2015MNRAS.454.2754Y}
requires    that the 
magnetization of the section of the pulsar wind producing the knot is low, $\sigma_w \leq 1$. Thus, the agreement of X-ray observation and low-sigma numerical models require that the equatorial part of the wind must have low magnetization, $\sigma_w \ll 1$.
The requirement of a low-magnetized equatorial part sides nicely with the results of PIC
modeling of Fermi-I acceleration in a weakly magnetized pair plasma, see \S \ref{PICI}. Thus, low-sigma MHD models and results of PIC simulations require/reproduce the Crab X-ray torus.

\subsubsection{Component-I: insights from PIC simulations}
\label{PICI}
Component-I is attributed to systematic Fermi-I acceleration at the relativistic shock
that terminates the pulsar wind. In pulsar winds, if the rotational and magnetic axes of
the central pulsar are misaligned, the wind around the equatorial plane consists of toroidal stripes of {alternating} magnetic polarity, separated by current sheets of hot plasma. Such a ``striped wedge'' extends at latitudes $\lambda\leq \xi$, where $\xi=\arccos(\hat{\mu}\cdot\hat{\Omega})$ is the angle between magnetic and rotational axes of the pulsar (i.e., the pulsar inclination angle). For  higher latitudes, the wind carries a non-alternating magnetic field.

Assuming that the stripes survive until the termination shock \citep[][but see \citealt{cerutti_17}]{lyubarsky_kirk_01,kirk_sk_03}, we now describe the physics of particle acceleration if the pre-shock flow carries a strong magnetic field of intensity $B_0$, oriented perpendicular to the shock normal and alternating with wavelength $\ell =c\,P$, where $P$ is the pulsar period. For the Crab nebula, we assume that the pre-shock  (before the alternating stripes get annihilated)  Lorentz factor of the wind is $\gamma_w\sim 100$, and the pre-shock magnetization is  $\sigma_w=B_0^2/(4 \pi \gamma_w \rho_0 c^2)\sim 3\times10^3$ (here, $B_0$ and $\rho_0$ are the wind magnetic field and mass density as measured in the frame of the nebula).
Importantly, if the magnetic energy is completely transferred to the particles (either in the wind or at the shock), the mean particle Lorentz factor increases from $\gamma_w$ up to 
\be
\gamma_p=\gamma_w\sigma_w \, .
\label{gammap}
\ee

 Although the magnetic field strength in the  wind is always $B_0$, the wavelength-averaged field $\langle B_\phi\rangle_\ell$ can vary from zero up to $B_0$, depending on the relative widths of the regions of  positive and negative field.
  In  pulsar winds, one expects $\langle B_\phi\rangle_\ell=0$ only in the equatorial plane (where the stripes are symmetric), whereas $|\langle B_\phi\rangle_\ell|/B_0\rightarrow1$ at high latitudes (more specifically, at latitudes  $\lambda\rightarrow\xi$).
As a proxy for latitude, we choose $\alpha_\ell=2\langle B_\phi\rangle_\ell/(B_0+|\langle B_\phi\rangle_\ell|)$, which varies between zero (at the midplane) and unity (at  $\lambda=\xi$). We now describe the expected shape of the particle spectrum as a function of $\alpha_\ell$ (or equivalently, latitude), assuming that both $\gamma_w$ and $\sigma_w$ are independent of latitude. Our arguments are motivated by the results of PIC simulations described in \cite{2011ApJ...726...75S}.

At the termination shock, the compression of the flow forces the annihilation of nearby field lines through driven magnetic reconnection \citep{lyubarsky_03, 2011ApJ...741...39S,sironi_spitkovsky_12}. The efficiency of field dissipation is a function of latitude.
 Taking $\langle\gamma\rangle/\gamma_w\gtrsim 0.8$ as a criterion for efficient dissipation, we find that for $\alpha_\ell\lesssim 0.5$ most of the magnetic energy is dissipated, and as a result of reconnection the mean particle Lorentz factor increases from $\sim\gamma_w$ up to $\gamma_p \sim\gamma_w\sigma_w$. 
The resulting shape of the reconnection-powered particle spectrum depends on the ratio $\ell/r_{L,hot}$, where $r_{L,hot} = \sqrt{\sigma_w}\,c/\omega_p$ (being $\omega_p$ the plasma frequency), namely the stripe wavelength measured in units of the {post-shock} particle Larmor radius (i.e., after dissipation has taken place, and the mean particle energy has increased by a factor of $\sigma_w$). For Crab parameters, this ratio is small, and the resulting reconnection-powered particle spectrum should resemble a broad Maxwellian  peaking at $\sim\gamma_w\sigma_w$ \citep{2011ApJ...741...39S}. 
A Maxwellian peaking at $\sim\gamma_w\sigma_w$ lying in a region with low residual magnetization is indeed invoked to explain the radiation spectrum of the Inner Knot.

At higher latitudes ($\alpha_\ell\gtrsim 0.5$) only a moderate or minor fraction of the
incoming field energy is dissipated.  For $0.5\lesssim\alpha_\ell<1$, the post-shock
particle spectrum consists of   two Maxwellian  components. The low-energy peak comes from cold plasma
with mean Lorentz factor $\sim \gamma_w$, whereas the high-energy part is populated by
hot particles that gained energy from field dissipation, so that their mean Lorentz
factor is now $\sim \gamma_w \sigma_w$. With increasing latitude, the high-energy part gets depopulated at the expense of the low-energy part. For $\lambda\ge\xi$, i.e., at latitudes beyond the striped wedge, only the low-energy part peaking at $\sim\gamma_w$ survives. Here, the magnetic energy per particle is still large. Further downstream, in the bulk of the nebula, turbulent dissipation of the dominant magnetic energy embedded in this portion of the wind will give rise to Component-II.


As we have just discussed, on most of the shock surface, shock-driven reconnection is not expected to produce non-thermal particle distributions. Rather, depending on the latitude, in the immediate post-shock region one expects a Maxwellian peaking at $\sim \gamma_w$ or  at $\sim \gamma_w\sigma_w$ (or a combination of the two). 
 We do not expect to see a simple thermal bump in the integrated spectrum, though.
The  oblique post-shock flow has a large spread in Lorentz factors, and
a corresponding large spread in the internal temperatures, \citep{2011MNRAS.414.2017K}.
Also, these leptons were at the lower end of the injection
spectrum $\gamma_{min,I} \sim \gamma_p$, and on the time scale of the Nebula they cooled off into the optical.
Variations in B-field and the spread in post-shock temperatures (for oblique shocks) can erase a simple thermal distribution.

The equatorial wedge at $\alpha_\ell\lesssim 0.03-0.1$ represents an exception -- here a systematic Fermi process can efficiently operate, in analogy to what happens in weakly magnetized shocks. Since the stripes are quasi-symmetric, the highest energy particles accelerated by the reconnection electric field can escape ahead of the shock, and be injected into a Fermi-I acceleration cycle by ``surfing'' the alternating magnetic fields (effectively, the shock in the equatorial plane has zero stripe-averaged field, and it displays the same acceleration capabilities of a weakly magnetized or unmagnetized shock, \citealt{spitkovsky_08, 2013ApJ...771...54S}). In the post-shock spectrum, the shock-accelerated particles populate a power-law tail with slope $p\simeq2.3$ that extends beyond the Maxwellian at $\gamma_w\sigma_w$ produced by reconnection. In summary, Component-I can be produced via Fermi-I acceleration in the equatorial region (at latitudes $\tan \lambda/\tan \xi\lesssim 0.03-0.1$): its minimum Lorentz factor is $\sim \gamma_w\sigma_w$, and the slope of the non-thermal tail is $p\simeq2.3$.

\subsection{Component-II: the turbulent nebula}

\subsubsection{Overview} 

We propose that magnetized turbulence generates reconnecting current sheets of different sizes in the bulk of the Nebula. Particles are then accelerated by magnetic reconnection in the current layers and by scattering off turbulent fluctuations. We suggest that this is the origin of Component-II. For a distribution of reconnecting sheets, with different sizes, that accelerate particles to some spectral slope $p$, to different $\gamma_{max}$ (determined by the size of the reconnection region and the inflow velocity), the final average spectrum will be a combination of intrinsic $p$, distribution of $ \gamma_{max}$, and how many particles are accelerated in a sheet with given $ \gamma_{max}$. 
In addition, reconnection regions may occur in environments with different $\sigma_w$ -- there will be scaling with the relative contribution of different regions. Finally, particles can be further accelerated by stochastic interactions with turbulent fluctuations, producing a nontermal tail that extends well beyond the energy gained via magnetic reconnection alone (Comisso \& Sironi 2018).

There are requirements on the spectrum of Component-II that have to be kept into consideration.
Radio electrons (the low energy part of the distribution produced by Component-II)
have $p_{II} \approx 1.6$ (for the Crab). Then there will be a cooling break in the IR. On top of that, there should be another mild break, in order to extend the spectrum from radio to gamma-rays, see Fig. \ref{Crab-spectrum-pictures}. The break may come from combination of particles accelerated in regions with different $\sigma_w$ - if larger regions have on average smaller magnetization (which, in reconnection, leads to steeper slopes) then the higher energy part of the spectrum will be softer.

We expect that  
extra break of Component-II  is related to the mean energy per particle. Thus, it is  the same as $\gamma_{I, min}$ of the shock-accelerated Component-I (see below).

 \subsubsection{Component-II: the problem of $\gamma_{min,II}$}
 
 \cite{1984ApJ...283..694K,1984ApJ...283..710K} recognized two important problems with the radio
emitting electrons in the Crab: (i) the minimal \Lf\ for radio emitting electrons, $ \gamma_{min,II }  \leq 100$,   is well below the average energy per particle $\gamma_p \geq \sim 10^6$. 
 It is expected that the termination shock heats particles to the \Lf\ $\gamma_p$  and  further accelerates beyond that via Fermi-I process. Thus, Component-II requires $ \gamma_{min,II }$ well below the expected average energy per particles in the wind. 

We already outlined a
possible resolution of $\gamma_{min,II}$ in \S \ref{PICI}. The pulsar wind is  relativistic, $\gamma_w \gg 1$, and magnetically dominated, $\sigma_w \gg 1$. Thus, the total energy per particle (in terms of $m_e c^2$) is $\gamma_p \sim \gamma_w \sigma_w$. Within the striped part of the wind this total energy is given to the particles.
 We assume $ \gamma_w \sim 10^2$ and $\sigma_w \sim 10^3$; thus,  $\gamma_{min,I} \sim \gamma_p \sim 10^5$. 
At the intermediate attitudes, where the wind is not striped, only the bulk energy is thermalized, giving  $\gamma_{min,II} \sim  \gamma_w \sim 10^2$.

\subsubsection{Turbulence is required to resolve the $\sigma$-problem}

\cite{1984ApJ...283..694K} identify the so-called sigma-problem of pulsar winds: models of pulsar magnetospheres predict $\sigma_w \gg 1$, while such flows cannot be accommodated with the non-relativistically expanding nebula.
Numerical simulations of the global structure of the Crab Nebula
\citep{2014MNRAS.438..278P}
observations of larger scale
structure in radio \citep{2017ApJ...840...82D}, as well as modeling of emission \citep{2016MNRAS.460.4135P} demonstrate that the structure of the Crab Nebula is highly turbulent.
This runs contrary to the \cite{1984ApJ...283..694K} model that envisions smooth evolution of the flow and the \Bf\ from the termination shock to the edges. Turbulent redistribution  of the magnetic flux from large scales to small scales, accompanied by the dissipation of magnetic energy, is required to resolve the sigma-problem \citep{2006NJPh....8..119L,2010MNRAS.405.1809L,2013MNRAS.431L..48P}.

\subsubsection{Very hard spectral indices in radio - not Fermi-I at shocks}
The majority of PWNe have radio spectral indices $\alpha \sim 0.1-0.2$ \citep{2014BASI...42...47G}. This implies a particle index $p \approx 1.2$.  Such hard spectra are not expected from a Fermi-I shock process, which typically produces $p> 2$ \citep[\eg][and \S \ref{PICI}]{1987PhR...154....1B}. On the other hand, reconnection in highly magnetized plasmas can produce spectra that approach $p \sim 1$:  in two dimensions \citep{zenitani_01,zenitani_07,jaroschek_04,bessho_05,bessho_07,bessho_10,bessho_12,hesse_zenitani_07,daughton_07,lyubarsky_liverts_08,cerutti_12b,2014ApJ...783L..21S,guo_14,guo_15a,liu_15,nalewajko_15,sironi_15,sironi_16,werner_16,kagan_16,kagan_18,petropoulou_18,2014ApJ...782..104C,2017JPlPh..83f6301L,2018JPlPh..84b6301L,2017JPlPh..83f6302L,2012ApJ...746..148C, 2015ApJ...806..167G,2015ApJ...815..101N,2016ApJ...816L...8W,2008ApJ...677..530Z},  and  in three dimensions \citep{zenitani_05b,2008ApJ...677..530Z,yin_08,liu_11,2011ApJ...741...39S, sironi_spitkovsky_12,kagan_13,cerutti_13b,2014ApJ...783L..21S,guo_15a,werner_17}.
There is a clear trend: higher sigma plasmas produce harder spectra, seeming to approach $p \sim 1$ in the very high-sigma limit.

\subsubsection{Component-II: insights from PIC simulations}
\label{CompoII}

The post-shock high-latitude flow has a small mean Lorentz factor $\gamma_w$, and most of
the energy per particle still resides in the magnetic field. When transported through the
nebula, the magnetic energy might be dissipated, as a result of reconnection in a
turbulent environment. It is well known that reconnection in magnetically dominated pair
plasmas can produce hard particle spectra
\citep{zenitani_01,2014ApJ...783L..21S,guo_14,werner_16}. Simulations of reconnection
have shown that $p\lesssim 2$ for $\sigma_w\gtrsim10$ (but see \citealt{petropoulou_18}),
and the slope tends asymptotically to $p\rightarrow1$ in the limit $\sigma_w\gg1$ appropriate for the Crab Nebula.  However, it is conceivable to assume that the superposition of spectra from different regions in the nebula will soften the space-averaged electron spectrum of Component-II to the observed $p \approx 1.6$. It remains to be explored whether such a superposition is a natural result of inhomogeneities in the structure of the nebula.



We thus envision that reconnection in the nebula can naturally give the hard spectrum required for the radio electrons, above $\sim\gamma_w$. In simulations of $\sigma_w\gg1$ reconnection of an isolated current sheet, the high-energy cutoff of the particle distribution evolves quickly up to $\gamma_{max} \sim 4 \gamma_w\sigma_w$ \citep{werner_16}, followed by a slower evolution to higher energies \citep{petropoulou_18}. In a turbulent environment, one might argue that particles energized by reconnection to Lorentz factors $\gtrsim \gamma_p=\gamma_w\sigma_w$ would be further accelerated by turbulent fluctuations (here, we neglect the factor of 4 between $\gamma_{max}$ and $\gamma_p$); in contrast, the particle spectrum between $\gamma_w$ and $\sim \gamma_p$ will be entirely determined by reconnection, since particles will be trapped in reconnection plasmoids and cannot be effectively scattered by turbulence.

Particle acceleration in a self-similar sequence of current sheets produced by a
turbulent cascade has recently been investigated by \cite{zhdankin_17} and \cite{2018PhRvL.121y5101C}. These works have demonstrated that the generation of a power-law particle energy spectrum extending beyond the mean energy per particle $\sim \gamma_w\sigma_w$ is a generic by-product of magnetized turbulence. More specifically, \cite{2018PhRvL.121y5101C} have shown that in a magnetized turbulent environment where turbulence and reconnection interplay, particle injection from the thermal bath at $\gamma_w$ up to $\sim \gamma_w\sigma_w$ is governed by the reconnection electric field; some particles are then further accelerated by stochastic interactions with turbulent fluctuations, producing a non-thermal tail extending much beyond the mean energy per particle. 
The power-law slope of the high-energy tail is harder for higher magnetizations and stronger turbulence levels. While the regime $\sigma_w\gg1$ appropriate for the Crab nebula is yet to be fully explored, we find that the slope can become harder than $p = 2$ \citep[][]{2018PhRvL.121y5101C}, but it is in general steeper than
  the slope generated by reconnection alone with the same parameters. 
  
  We then argue that a break in the particle distribution of Component II should appear around $\gamma_p\sim \gamma_w\sigma_w$ (this is also the location of the injection Lorentz factor for Component I): at lower energies, energization is controlled by reconnection; at higher energies, reconnection-injected particles are further accelerated by turbulence. We remark, though, that at this point this is merely an hypothesis motivated by the distinct roles of the two acceleration mechanisms, as discussed in \citet{2018PhRvL.121y5101C}. To assess the validity of this assumption, one will need simulations of turbulence with $\sigma_w\gg1$ 
  and a large dynamic range between the energy-carrying turbulent scales and the Larmor radius of particles energized by reconnection.

In summary, we expect the particle spectrum of Component-II to be a broken power-law: from $\sim\gamma_w$ to $\sim \gamma_w\sigma_w$ the space-averaged spectrum has slope $p \approx 1.6$ (but it is locally expected to be as hard as $p=1$), in response to acceleration via reconnection; beyond the break at $\sim \gamma_w\sigma_w$, particles are accelerated by a stochastic process akin to Fermi II, with a steeper slope $p\gtrsim 2$.

The timescale of the stochastic acceleration process is comparable to that of magnetic reconnection in the strong turbulence scenario considered here. Indeed, in this regime, the energy diffusion coefficient is $D_{\gamma\gamma} \sim 0.1 \gamma^2 (c/l)$, as measured in  \cite{2019arXiv190103439W} and Comisso and Sironi 2019 (in prep.). Here, $l$ indicates the integral scale of turbulence, which specifies the energy-containing scale, and is also indicative of the size of the longest current sheets in the turbulent environment. Therefore, the stochastic acceleration timescale $t_{acc} \sim \gamma^2/D_{\gamma\gamma} \sim 10 l/c$ is fast and comparable to that of magnetic reconnection.

We expect that at the largest scales the statistics of current sheets might not be described as a self-similar sequence controlled by turbulent motions. There could be non-self-similar 
current sheets that result from large scale collisions of flux tubes \citep{2017JPlPh..83f6302L}. In this case, if $\sigma_w\sim 10-100$, the  reconnection might be responsible for particle acceleration up to the maximum available potential (for such magnetizations, $p\sim 2$, and the maximum particle energy is not constrained by the fixed energy content of the system). Particle acceleration at these large-scale current sheets, Component-II,  can extend up to the synchrotron burn-off limit and beyond, thus powering the  Crab Nebula gamma-ray flares \citep{2017JPlPh..83f6301L,2018JPlPh..84b6301L}

\section{Discussion}

In this paper we advance a two-component model of the Crab Nebula broad-band  synchrotron
emission. Component-I arises from the shock acceleration at the equatorial part of the
pulsar wind termination shock (and thermalization at higher latitudes), while
Component-II is generated from the interplay between reconnection events and turbulent
fluctuations in the bulk of the Nebula.  We suggest that Component-II extends from radio
to gamma-rays, but is subdominant to Component-I in the range from the optical band to
hard X-rays. 
There are two independent motivations for the separate  Component-II related to magnetic
reconnection and turbulence: very hard radio spectra and gamma-ray flares -- both are inconsistent with shock acceleration.

We envision the following structure/ acceleration properties of the Crab PWN, Fig. \ref{PWN-overview}
\begin{itemize}
\item The Crab pulsar produces a relativistic highly magnetized wind.
\item The equatorial part of the termination shock, of the order of $\sim 10^\circ$ around the equator, 
accelerates X-ray emitting particles via the reconnection-mediated  Fermi-I
mechanism, while  at higher latitudes the reconnection-mediated shock  just
thermalizes the wind. This produces Component-I with $\gamma_{min,I} \sim \gamma_w \sigma_w \sim 10^5$. 
\item Component-I  is in the fast cooling
regime on the time sale of the Nebula. Cooled leptons produce optical and near-IR emission (with a spectral index close to $\alpha =0.5$, steepening towards the edges).
\item At even higher latitudes the termination shock thermalizes  only the bulk energy of the wind, $\gamma_{min,II} \sim \gamma_w\sim 10^2$.
\item  Turbulence in  high-sigma regions in the mid-to-high latitudes leads to the formation of reconnecting current sheets and turbulent fluctuations that accelerate  particles from radio to $\sim 100$ MeV. Rare large-scale reconnection events lead to Crab gamma-ray flares. Radio and gamma-ray flares are produced by the same population (Component-II), different from X-rays (Component-I)
\item The above-GeV photons are produced by IC scattering of the synchrotron emission, in a  conventional way.  Subdominant extension of the synchrotron emission of Component-II to MeV energies will not change appreciably the IC component which has been previously calculated using the observed synchrotron spectra \citep{1996MNRAS.278..525A,2010A&A...523A...2M,2012MNRAS.427..415M}
\end{itemize}

We interpret the UV peak in the spectrum of the Crab Nebula as arising from the peak
injection of particles accelerated at the equatorial part of the termination shock
with a power-law index $p_I \approx 2.2$. The UV peak energy corresponds to the $\gamma_p\sim \gamma_w \sigma_w \sim $ few $\times 10^5$.
Optical and near-IR photons are produced by cooled leptons, both those accelerated at the equatorial part of the termination shock and those that are just thermalized at the shock at intermediate latitudes. This naturally produces spectral index $\sim 0.5$.

 Radio emitting particles require injection index $p_{II} \approx 1.6$, which is too
hard for the Fermi-I acceleration. We suggest that radio emitting particles are
accelerated in turbulence mediated reconnection events. Component-II experiences two
breaks: a cooling break in the IR
and an intrinsic injection break corresponding to the minimal \Lf\ of Component-I. 
In the optical and UV Component-II is mostly
overwhelmed by Component-I.
The injection break in Component-II above the IR is needed for Component-II not to overshoot Component-I up to hard X-rays (if there were no break in Component-II, it would produce 
a slightly rising spectrum $\nu F_\nu \propto \nu^{0.2}$ above the cooling break, that may overshoot the falling $\nu F_\nu \propto \nu^{-0.1}$ of Component-I in the X-rays).
 We hypothesize that  Component-II extends to the synchrotron burning off limit of
100 MeV. In our model, magnetized turbulence with reconnecting current sheets accelerates both the radio electron
and also produces the Crab $\gamma$-ray flares. 

Importantly, the implied bulk \Lf\ in the wind is fairly low, $\gamma_w \sim 10^2$. Conventional models of particle acceleration near the polar caps \citep[\eg][]{AronsScharlemann} predict $\gamma_w \sim 10^4-10^6$. On the other hand, there is still no complete kinetic model of pulsar \mss\ that include pair production and particle acceleration self-consistently. 

In the present model the region $\sim 0.01-1$ eV is the most complicated: this is where several competing effects contribute nearly equally: (i) this is where the overall cooling break is for both components ($\sim 0.01$ eV); (ii) where the peak of Component-I is ($\sim 1$ eV); (iii) where  Component-I cools into optical, $1 {\rm eV}\rightarrow 0.1 {\rm eV}$. All these estimates are  a strong function of local B-field, hence large variations are expected.

 The model has several predictions:
 \begin{itemize}
 \item {\it Details of the spectrum of the torus: shape of the UV peak.} The shock-produced Component-I is in the fast cooling regime on the lifetime of the Nebula, but is in the slowly cooling regime on the life-time of particles in the torus (years). In our model the UV peak is the injection break of Component-I. Thus, the spectrum of the torus should change from 
 $\nu F_\nu \propto \nu^{4/3}$ below the peak to $\nu F_\nu \propto \nu^{(3-p_I)/2}
\propto \nu^{0.4} $ (hence, it is a break not a peak). Only at higher frequencies should
it steepen to $\nu F_\nu \propto \nu^{(2-p_I)/2} \propto \nu^{-0.1} $. (These are asymptotic estimates, for well separated cooling, injection and the observed frequencies; more detailed computations are needed to address the shape of the UV bump.) There is another testable prediction of the model related to the UV peak. If higher latitudes only thermalize the flow, this may have an effect on the shape of the peak. 
 \item {\it Possible reappearance of Component-II above hard X-rays.} In the model, Component-II becomes
subdominant to Component-I above the cooling break at $\sim 0.01$ eV. It may also experience an injection break at $\gamma_{I, min}$. Depending on the value of the
spectral index $p_{II,above}$ above the break, Component-II may eventually dominate over 
Component-I (which is slowly
decreasing with frequency above the peak). The MeV COMPTEL bump, \S \ref{COMPTEL}, offers tantalizing hints.
  \item {\it Flares in the low MeV range.} In our model the MeV COMPTEL bump, \S
\ref{COMPTEL}, is interpreted as a different component from the lower energy X-rays; it
extends to GeV energies and occasionally produces $\gamma$-ray flares. We may expect
flares in the lower MeV range.  Flares from Component-II also occur in the X-ray region, but at those energies    the integrated flux is dominated by Component-I, so that only the brightest X-ray flares could possibly be detectable.
   \item We expect  a Doppler-broadened thermal peak in the UV, corresponding to just thermalized wind component at high latitudes. 
 \end{itemize}

There are a number of issues that need to be addressed within the model, that we leave to the future study:  a combined radiative model of the two synchrotron emitting components; (ii) calculations of the SSC spectrum within the model (\eg, whether the spatially-dependent \Bf\ of $\sim 500\mu$G is consistent with the IC signal; (iii) calculations of the spectral evolution of Component-I (\eg, one would predict temporary  hardening in optical as the peak particles from Component-I first cool into optical from the UV into optical and then into IR).

\section*{Acknowledgments}
This work had been supported: ML by NSF grant AST-1306672, DoE grant DE-SC0016369 and
NASA grant 80NSSC17K0757;  LS:  DoE DE-SC0016542, NASA Fermi NNX-16AR75G, NASA ATP
NNX-17AG21G, NSF ACI-1657507, and NSF AST-1716567; PS: NASA contract NAS8-03060.

ML would like to thank Laboratoire
Univers et Particules de l'Universite de Montpellier for hospitality.

We would like to thank Fabio Acero, Joe DePasquale, Gloria Dubner,  Yves Gallant, Elisabeth Jourdain, Oleg Kargaltsev, Lucien Kuiper,  George Pavlov, Mark Pearce, Matthieu Renaud, Steve Reynolds, Jean-Pierre Roques and Jacco Vink for comments and discussions.

\bibliographystyle{apj}

 \bibliography{/Users/maxim/Home/Research/BibTex}

\end{document}